\documentclass[11pt,fullpage]{article}

\usepackage{amsmath}
\usepackage{amsfonts}
\usepackage{algorithm}
\usepackage{algpseudocode}
\usepackage{graphicx}
\usepackage{subcaption}
\usepackage{enumitem}
\usepackage{comment}
\usepackage{fullpage}
\usepackage[obeyspaces,spaces]{url}
\usepackage{hyperref}
\usepackage{times}

\begin{document}

\date{October 2021}

\title{Adaptive Elastic Training for Sparse Deep Learning on Heterogeneous Multi-GPU Servers}

\author{Yujing Ma$^{1}$, Florin Rusu$^{1,2}$, Kesheng Wu$^{2}$, Alexander Sim$^{2}$\\
\{yma33,frusu\}@ucmerced.edu, \{kwu,asim\}@lbl.gov\\
$^{1}$University of California Merced\\
$^{2}$Lawrence Berkeley National Laboratory\\
}

\maketitle

%%%%%%%%%%%%%%%%%%%%%%%%%%%%%%%%%%%%%%%%%%%%%%%%%%%%%%%%%%%%%%%%%
%\input{abstract}
\begin{abstract}

Motivated by extreme multi-label classification applications, we consider training deep learning models over sparse data in multi-GPU servers. The variance in the number of non-zero features across training batches and the intrinsic GPU heterogeneity combine to limit accuracy and increase the time to convergence. We address these challenges with Adaptive SGD, an adaptive elastic model averaging stochastic gradient descent algorithm for heterogeneous multi-GPUs that is characterized by dynamic scheduling, adaptive batch size scaling, and normalized model merging. Instead of statically partitioning batches to GPUs, batches are routed based on the relative processing speed. Batch size scaling assigns larger batches to the faster GPUs and smaller batches to the slower ones, with the goal to arrive at a steady state in which all the GPUs perform the same number of model updates. Normalized model merging computes optimal weights for every GPU based on the assigned batches such that the combined model achieves better accuracy. We show experimentally that Adaptive SGD outperforms four state-of-the-art solutions in time-to-accuracy and is scalable with the number of GPUs.

\end{abstract}

%%%%%%%%%%%%%%%%%%%%%%%%%%%%%%%%%%%%%%%%%%%%%%%%%%%%%%%%%%%%%%%%%
%\input{introduction}
\section{INTRODUCTION}\label{sec:intro}

Deep learning models have been shown to achieve high accuracy for many classification problems across diverse application domains, including speech recognition~\cite{speech-sgd}, finance~\cite{finance-sgd}, and renewable energy~\cite{combustion-dnn}. However, this is heavily dependent on the amount of training data and the size of the model. As these two values increase, building accurate deep learning models becomes time-consuming even on specialized hardware accelerators such as GPUs~\cite{facebook:large-batches} and TPUs~\cite{imagenet-minutes:iclr-2020} because of their reduced memory---which requires many slow data transfers with the CPU~\cite{slide:arxiv}. Thus, in order to scale training, parallel processing across multiple GPUs~\cite{geeps,crossbow} becomes a necessity. This approach is facilitated by the preponderance of multi-GPU computing architectures both on supercomputers~\cite{perlmutter} and in the cloud~\cite{aws-ec2-gpu}.

%%%%%%%%%%%%%%%%%%%%%%%%%%%%%%%%%%%%%%
\paragraph*{Extreme Multi-label Classification (XML)}
We consider extreme multi-label classification as a motivating application for our work. The objective in XML is to tag a data point with the most relevant subset of labels from an extremely large set that can contain up to millions of possible labels. The Extreme Classification Repository~\cite{Xclassification-dataset-repo} contains an exhaustive collection of real datasets, code, and experimental results for XML algorithms. In a typical dataset (Table~\ref{tbl:dataset}), the feature vector, as well as the associated labels, are highly sparse, thus processing requires sparse linear algebra operations. This is quite different from the image classification benchmarks -- which are based on dense linear algebra -- used almost exclusively to evaluate the existing multi-GPU training algorithms~\cite{facebook:large-batches,imagenet-minutes:icpp-2018,crossbow}. AttentionXML~\cite{attention-xml}, DeepXML~\cite{deep-xml}, and LightXML~\cite{light-xml} are some of the most recent deep learning XML algorithms that achieve the highest accuracy. Although these algorithms use complex learning architectures, such as the transformer model, tree-based models, and generative cooperative networks, they achieve top-1 accuracy below 50\% even after training for hours on multi-GPU servers with two or more GPUs.

%%%%%%%%%%%%%%%%%%%%%%%%%%%%%%%%%%%%%%
\begin{figure}
\centering
	\includegraphics[width=.9\linewidth]{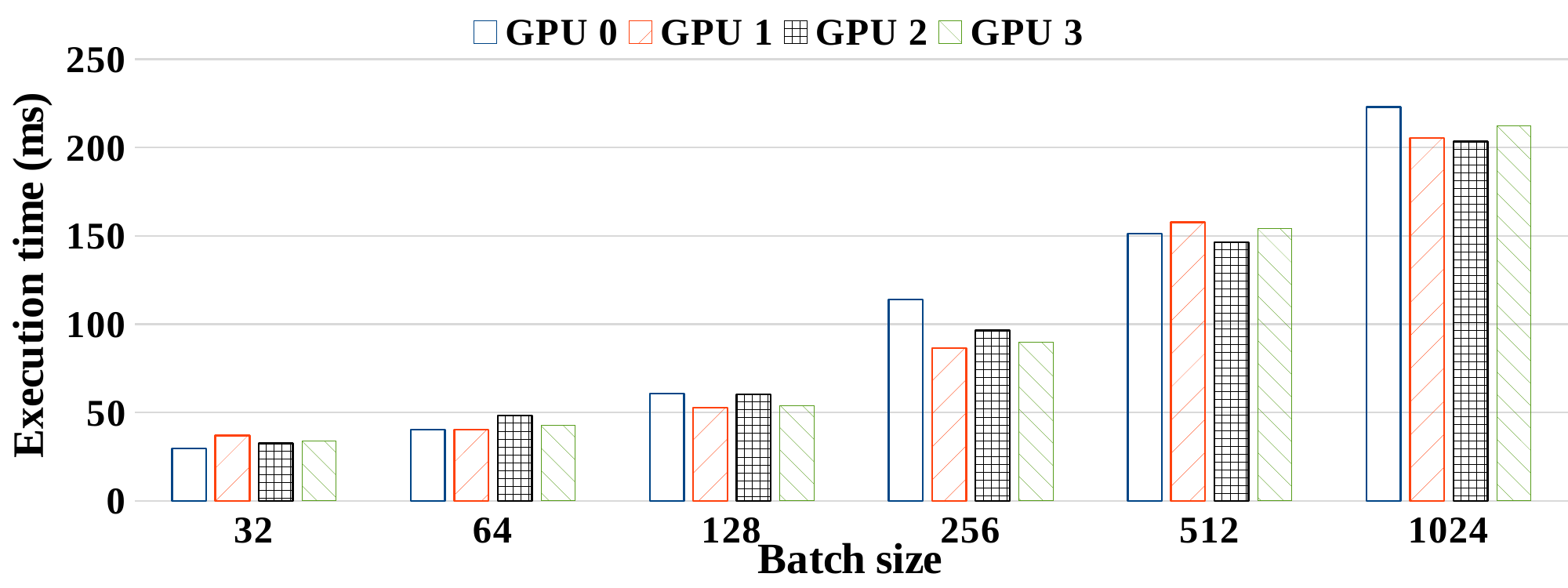}
\caption{Multi-GPU heterogeneity on training a deep learning model with an identical batch of sparse data.}
\label{fig:gpu-speed}
\end{figure}
%%%%%%%%%%%%%%%%%%%%%%%%%%%%%%%%%%%%%%

%%%%%%%%%%%%%%%%%%%%%%%%%%%%%%%%%%%%%%
\paragraph*{Heterogeneity in Multi-GPU Architectures}
We argue that an important reason for the high XML training time is the uniform handling of multiple GPUs. There are two major sources of heterogeneity in a multi-GPU architecture for training on sparse data. The first source is due to the differences among identical GPUs. The clock rate and memory latency display oscillations on GPUs with the same model from the same vendor. This effect is amplified when multiple GPUs are integrated on the same server, in which case the execution time varies within observable ranges. For example, given the same training batch, the gap between the fastest and slowest GPU is as large as 32\% when performing an epoch of the training algorithm on a server with 4 NVIDIA V100 GPUs (Figure~\ref{fig:gpu-speed}). The second source of heterogeneity is specific to sparse data. The number of non-zero features varies significantly among the training samples. When we partition training data into batches, it is almost impossible to guarantee that every batch contains the same number of non-zeros. Since sparse linear algebra operations are sensitive to the cardinality of their input, the effect is variation in processing across batches. The combination of these two factors results in significant differences among GPUs, which impacts negatively the time-to-accuracy of the training algorithm.

%%%%%%%%%%%%%%%%%%%%%%%%%%%%%%%%%%%%%%
\paragraph*{Problem}
We investigate multi-GPU training algorithms for deep learning. We consider two dimensions of this problem that have not been adequately addressed before. We target \textit{sparse training data} with sparse labels having high dimensionality as found in XML classification tasks. The second dimension is the \textit{heterogeneity} of the environment consisting of GPUs that exhibit significant variance in executing identical tasks. When data sparsity and GPU heterogeneity are combined, the delay among GPUs at the synchronization barriers becomes a severe bottleneck that can hinder time-to-accuracy dramatically. Solutions focused exclusively on determining the optimal synchronization frequency are challenging since they depend both on the data/problem and the relative GPU performance. Instead we target \textit{adaptive algorithms} that monitor the execution continuously and configure the workload for every GPU according to its relative performance. However, this introduces variation across the local models trained by different GPUs, which requires the design of an appropriate model merging scheme.

%%%%%%%%%%%%%%%%%%%%%%%%%%%%%%%%%%%%%%
\paragraph*{Contributions}
We introduce Adaptive SGD -- an adaptive elastic model averaging stochastic gradient descent algorithm for heterogeneous multi-GPUs -- characterized by dynamic scheduling, adaptive batch size scaling, and normalized model merging.
\begin{itemize}[leftmargin=*,noitemsep,nolistsep]
\item Instead of statically assigning batches to GPUs, in dynamic scheduling batches are allocated based on the relative GPU processing speed. This process is controlled by fixing the number of training samples processed between two model merging stages.

\item Since execution-driven allocation can lead to a different number of model updates across GPUs -- which is problematic for accuracy -- batch size scaling assigns larger batches to the faster GPUs and smaller batches to the slower ones, with the goal to arrive at a steady state in which all the GPUs perform the same number of model updates. The batch sizes are continuously updated following a linear function that quantifies the deviation from the expected number of updates, while guaranteeing a minimum degree of GPU utilization and imposing strict bounds on model replica staleness.

\item Normalized model merging computes optimal weights for every GPU based on the assigned batches. The underlying principle is to prioritize the replicas updated more frequently and -- secondarily -- with gradients derived from larger batch sizes. In order to increase the importance of the relevant model replicas, perturbation is added to the normalized weights when the replicas are well-regularized.
\end{itemize}

\smallskip
\noindent
We implement Adaptive SGD in the HeteroGPU framework~\cite{hetero-gpu:github} for sparse deep learning on multiple GPUs and compare its performance against four alternative methods. Due to the careful handling of heterogeneity -- which allows a more thorough exploration of the optimization space -- Adaptive SGD outperforms all the competitors in time-to-accuracy. In fact, Adaptive SGD always achieves the highest accuracy among all the algorithms. Moreover, as we increase the number of GPUs, Adaptive SGD exhibits both faster time-to-accuracy and less epochs to achieve an accuracy target. This confirms its superior scalability.

%%%%%%%%%%%%%%%%%%%%%%%%%%%%%%%%%%%%%%
\paragraph*{Outline}
The paper is organized as follows. Preliminaries on parallel training of deep learning models and related work are presented in Section~\ref{sec:prelims}. The Adaptive SGD algorithm is introduced in Section~\ref{sec:adap-elastic-sgd}. Implementation considerations are discussed in Section~\ref{sec:implementation}, while the extensive experimental evaluation is presented in Section~\ref{sec:experiments}. The conclusions are synthesized in Section~\ref{sec:conclusions}.

%%%%%%%%%%%%%%%%%%%%%%%%%%%%%%%%%%%%%%%%%%%%%%%%%%%%%%%%%%%%%%%%%
%\input{preliminaries}
\section{PRELIMINARIES}\label{sec:prelims}

In this section, we first introduce the basics of deep learning and then provide a detailed discussion on training models across multiple GPUs. We conclude with a thorough analysis of the related work.

%%%%%%%%%%%%%%%%%%%%%%%%%%%%%%%%%%%%%%%%%%%%%%%%%%%%%%%%%%%%%%%%%
\subsection{Deep Learning}\label{ssec:deep-learn}

The central component of a deep learning algorithm is a \textit{Deep Neural Network (DNN)}~\cite{deep-nets-sgd}. A DNN is a layered network that takes as input an example/sample given as a feature vector and produces the probability this example belongs to a certain class from a predefined set. The intermediate layers of the DNN are called hidden and they represent the model to be learned. This is achieved through supervised training over a dataset that includes the corresponding label -- ground truth -- for every sample. 

%%%%%%%%%%%%%%%%%%%%%%%%%%%%%%%%%%%%%%
\paragraph*{Stochastic Gradient Descent (SGD)}
SGD is the most common method to train DNN models~\cite{bottou:optim-methods-scale-ml}. At high-level, SGD iteratively computes the gradient of the loss function -- which is a measure of the error between the predicted and the true labels -- over the training dataset and moves the model in the opposite direction of the gradient---which results in a decrease of the error. In practice, the gradient is only estimated from a random \textit{batch of training samples} in order to reduce the number of accessed samples. SGD requires three passes over the DNN. In the forward pass, the predicted labels are computed based on the current model. The backward pass implements the chain rule of calculus for computing the gradient of a composite function starting from the predicted labels. The third pass -- which is also forward -- updates the model with the computed gradient using the \textit{learning rate} hyperparameter. SGD can be stopped either after a fixed number of iterations, i.e., \textit{epochs}, or when there is no significant drop in the error.

\medskip
\noindent
The performance of a DNN model is assessed by measuring its error/accuracy in predicting the labels of a test dataset that is not used in training. This measure is known as \textit{test accuracy}. The goal of training is to reduce the time to reach a target level of test accuracy, i.e., \textit{time-to-accuracy}. Two factors determine the time-to-accuracy. The first is the number of epochs required by SGD, known as \textit{statistical efficiency}, while the second factor is the execution time of an epoch---known as \textit{hardware efficiency}. These two factors are themselves dependent on the hyperparameters -- batch size and learning rate -- of the SGD algorithm.

%%%%%%%%%%%%%%%%%%%%%%%%%%%%%%%%%%%%%%%%%%%%%%%%%%%%%%%%%%%%%%%%%
\subsection{Multi-GPU SGD Training}\label{ssec:multi-gpu}

The standard approach to parallelize SGD is to partition the samples across the available GPUs and allocate a separate model replica to every GPU. Model replica transfer and synchronization become the dominant challenges in this setting. Depending on how they are addressed, two parallel SGD strategies are possible.

%%%%%%%%%%%%%%%%%%%%%%%%%%%%%%%%%%%%%
\begin{figure}[htbp]
\centering
\includegraphics[width=\textwidth]{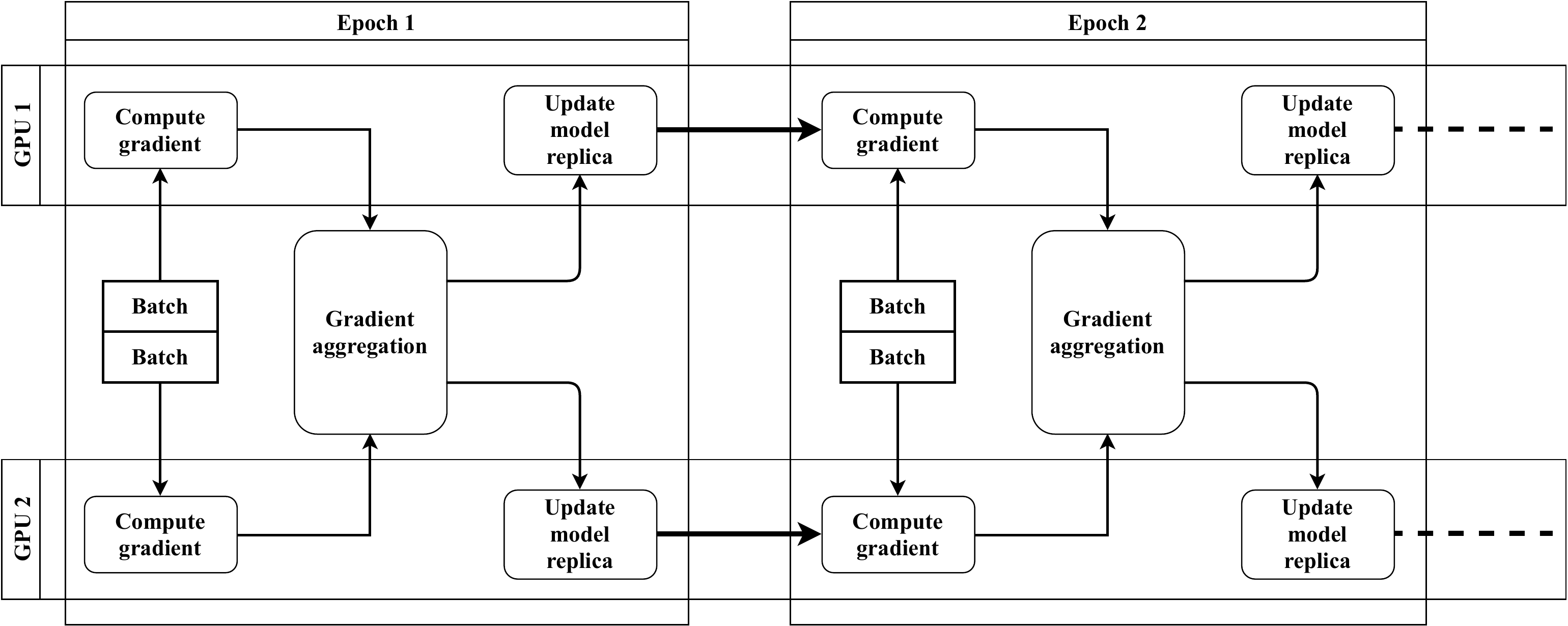}
\caption{Parallel synchronous SGD with gradient aggregation.}
\label{fig:gradient-agg}
\end{figure}
%%%%%%%%%%%%%%%%%%%%%%%%%%%%%%%%%%%%%

%%%%%%%%%%%%%%%%%%%%%%%%%%%%%%%%%%%%%%
\paragraph*{Gradient Aggregation}
Gradient aggregation -- or synchronous SGD~\cite{sync-vs-asynch-sgd} -- requires all the GPUs to have an identical replica of the model at every epoch. This guarantees that the same model is used to compute the gradient by every GPU and local gradients can be aggregated following a sound formula. Figure~\ref{fig:gradient-agg} depicts the gradient aggregation process for two GPUs. In an epoch, every GPU computes a partial gradient from a batch of training samples having the same size across all the GPUs and the identical model replica. The partial gradients are aggregated -- summed or averaged -- by coordinating all the GPUs. The aggregated gradient is shared among all the GPUs and applied to update every local model replica. Once this is done, the next epoch can begin. The synchronization imposed by gradient aggregation at every epoch is the main limitation of synchronous SGD---known as the straggler problem~\cite{crossbow}. Asynchronous SGD~\cite{hogwild,hogbatch,hogwild-disk,bgd-vs-sgd:ipdps-2019,lock-free-sgd:ipdps-2021} transforms gradient aggregation into a completely asynchronous process in which a GPU transitions to the next epoch immediately after its partial gradient is added to the aggregated gradient. The value accumulated thus far is used to update its local model replica. However, if performed over a large number of epochs, asynchronous SGD can result in divergent model replicas and poor convergence.

%%%%%%%%%%%%%%%%%%%%%%%%%%%%%%%%%%%%%
\begin{figure}[htbp]
\centering
\includegraphics[width=\textwidth]{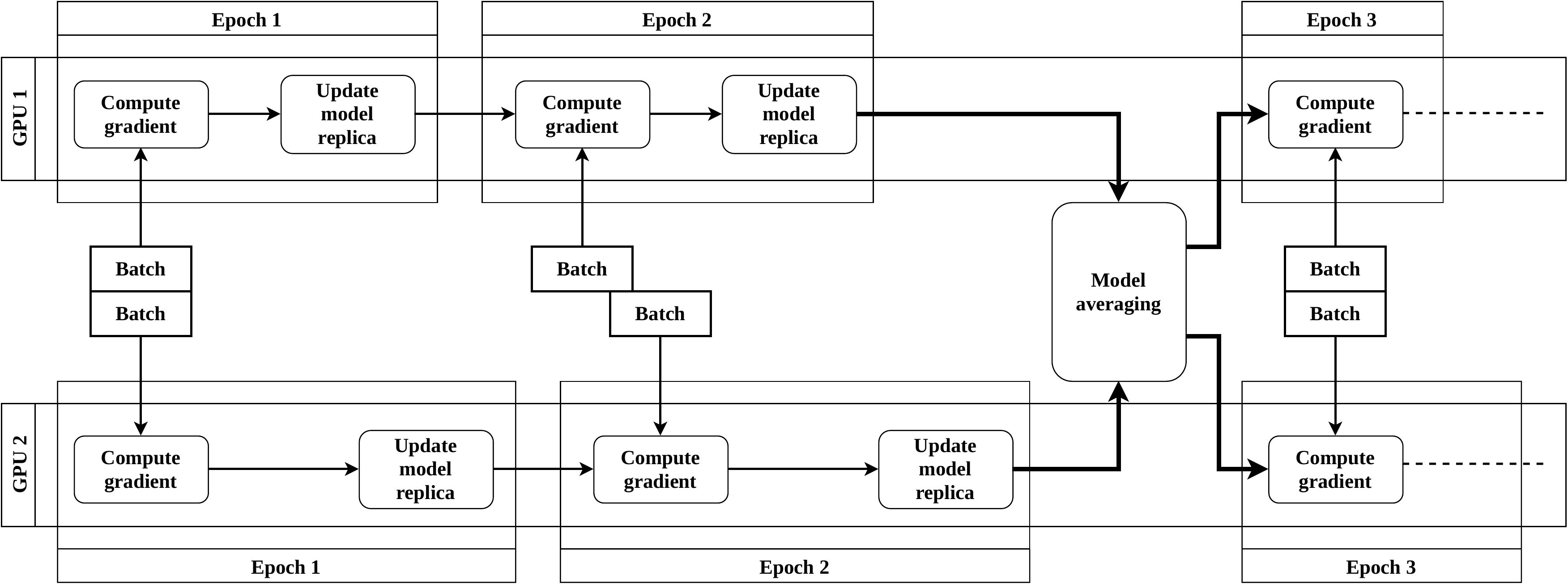}
\caption{Parallel SGD with elastic model averaging.}
\label{fig:elastic-model-avg}
\end{figure}
%%%%%%%%%%%%%%%%%%%%%%%%%%%%%%%%%%%%%

%%%%%%%%%%%%%%%%%%%%%%%%%%%%%%%%%%%%%%
\paragraph*{Elastic Model Averaging}
The elastic model averaging strategy~\cite{elastic-sgd} -- or K-step averaging~\cite{zhou2018convergence} -- is a tradeoff between synchronous and asynchronous SGD. It alleviates the burden in synchronous SGD by performing synchronization only after a predefined number of epochs---or batches. The model divergence issue from asynchronous SGD is addressed by averaging model replicas instead of gradients. By controlling the synchronization frequency, elastic model averaging generalizes gradient aggregation. It becomes synchronous SGD when aggregation is performed after every epoch---notice that gradient and model averaging are equivalent in this case. When the frequency is set to infinity and the local gradient/model replicas are references to a shared global gradient/model, elastic model averaging morphs into asynchronous SGD. Figure~\ref{fig:elastic-model-avg} depicts a visual illustration of the elastic model averaging workflow. In the example, we synchronize the model replicas on two GPUs after two sample batches. The GPUs progress independently until the model averaging stage, which is performed less often. While the synchronization cost is amortized, this does not eliminate the straggler problem since the delay among GPUs is cumulated across epochs. The delay depends both on the discrepancy between GPUs, as well as on data sparsity, and can vary from one epoch to another. Compared to Figure~\ref{fig:gradient-agg}, the improvement consists in one less gradient/model merging. In general, the reduction over gradient aggregation is inversely proportional to the synchronization frequency---the higher the frequency, the lower the reduction.

%%%%%%%%%%%%%%%%%%%%%%%%%%%%%%%%%%%%%%
\paragraph*{Analytical Comparison}
Although model averaging is a generalization of gradient aggregation, there is a subtle difference between them. The number of model updates per epoch is always one in gradient aggregation. In the case of model averaging, there is an update for every replica---two in Figure~\ref{fig:elastic-model-avg}. Thus, if we consider the batch size per GPU to be identical, the number of updates in model averaging is larger. However, the corresponding gradient is estimated from a smaller batch -- half the size in Figure~\ref{fig:elastic-model-avg} -- compared to gradient aggregation. The effect of these subtle differences is complex and depends on the values of the SGD algorithm hyperparameters. Although the relationship between the batch size and learning rate -- on one side -- and convergence (time-to-accuracy), number of updates (statistical efficiency), and utilization (hardware efficiency) -- on the other -- is well-known~\cite{imagenet-minutes:icpp-2018}, there is an ongoing debate about the optimal batch size -- small or large -- and learning rate~\cite{dont-decrease-lr-increase-batch,large-batch-empirical,revisit-small-batch}. Small batches generate more model updates, which have higher statistical efficiency. However, they do not saturate the high GPU throughput and result in low hardware efficiency. Overall, the combined effect on time-to-accuracy is not straightforward. A practical solution is to increase the batch size and the learning rate proportionally, i.e., linear scaling~\cite{one-weird-trick}. While this increases utilization and makes the steps taken by SGD both larger and more accurate, it also creates two problems---limiting the search space in the first epochs and introducing convergence instability close to the minimum~\cite{bertsekas:igd}. Fortunately, these can be addressed with warmup~\cite{facebook:large-batches} -- the learning rate is increased only after a certain number of epochs -- and adaptively increasing the batch size and learning rate across epochs. In~\cite{adaptive-batch-size-constant,dont-decrease-lr-increase-batch}, the batch size is increased by a pre-specified constant factor in each epoch, while in~\cite{adaptive-batch-size-variance:mp-2012,adaptive-batch-size-variance:aistats-2017} the batch size is dynamically adapted -- increased -- based on gradient variance estimates computed from previous batches. Furthermore, the variance is coupled with the learning rate in~\cite{couple-adaptive-batch-size-learn-rate} to provide an even more specialized criterion to select the optimal batch size. Based on these, given the larger batch size in gradient aggregation, a proportionally larger learning rate is required in order to achieve a similar effect to model averaging. However, beyond a certain point, a large learning rate impacts model convergence negatively~\cite{facebook:large-batches}. Thus, model averaging is a more reliable algorithm~\cite{crossbow}.

%%%%%%%%%%%%%%%%%%%%%%%%%%%%%%%%%%%%%%%%%%%%%%%%%%%%%%%%%%%%%%%%%
\subsection{Related Work}\label{ssec:rel-work}

%%%%%%%%%%%%%%%%%%%%%%%%%%%%%%%%%%%%%%
\paragraph*{Multi-GPU Training}
Out of the two strategies -- data and model partitioning -- to parallelize deep learning training over multiple GPUs~\cite{dnn-par+dis-survey}, only data partitioning is widely supported by existing systems. This is due to the more complicated communication pattern and pipelining required by efficient model partitioning. There are two data partitioning strategies in TensorFlow~\cite{tensorflow,tensorflow-web}, both of them implementations of gradient aggregation. In the mirrored strategy, there is a replica of the model on every GPU. The replicas are kept consistent by synchronously updating them with the aggregated gradient computed by an all-reduce operation implemented with the NVIDIA Collective Communication Library (NCCL)~\cite{nvidia-nccl} or with the Horovod~\cite{horovod,horovod-web} ring-all-reduce operation. In the central storage strategy, there is a single copy of the model stored on the CPU---which performs all the updates. As in the case of the mirrored strategy, the gradients are computed by the GPUs. The difference between mirrored and central storage is that all the communication is done among GPUs in mirrored, while data is transferred only between the CPU and every individual GPU in central storage. Variations of these strategies are supported by other deep learning systems, including MXNet~\cite{mxnet,mxnet-web}, CNTK~\cite{cntk,cntk-web}, and PyTorch~\cite{pytorch,pytorch-web}. Elastic model averaging is implemented only in CNTK. In CROSSBOW~\cite{crossbow}, every GPU has a local replica of the model, while the global model -- which is the average of the local replicas -- can be either mirrored across GPUs or centrally stored on CPU. Similar to elastic model averaging, the local replicas are allowed to evolve independently. However, they are corrected based on their divergence from the global model after every batch. The existence of two different models on every GPU doubles the memory requirement or, alternatively, the data transferred. None of these multi-GPU strategies consider variations in GPU processing and how to integrate them effectively in training---which is the purpose of this work.

%%%%%%%%%%%%%%%%%%%%%%%%%%%%%%%%%%%%%%
\paragraph*{Elastic Distributed Training}
The standard approach to scale deep learning training to distributed clusters is Parameter Server (PS)~\cite{parameter-server}, which is essentially the TensorFlow central storage strategy~\cite{tensorflow-web} enhanced with network communication. Given the larger latency, training data are statically assigned to the cluster nodes and the model traffic is kept as low as possible by applying elastic model averaging~\cite{elastic-sgd}. Asynchronous decentralized SGD~\cite{asynch-elastic-sgd} reduces communication traffic further by requiring a local replica to perform model averaging with only one other replica. The main difference in a single-server multi-GPU setting is that batch scheduling is dynamic. This allows GPUs to process a different number of training samples. Model averaging and synchronization are performed after a mega-batch -- a configurable number of training samples -- which makes merging more complicated than the learning rate driven stale synchronous PS~\cite{stale-synch-ps,hetero-ps} and the solution reported in~\cite{crossbow}, where the GPUs have the same batch size.

%%%%%%%%%%%%%%%%%%%%%%%%%%%%%%%%%%%%%%
\paragraph*{Adaptive Optimization}
Variance reduction algorithms~\cite{svrg,saga,flex-ps} aim to reduce the variance of stochastic gradients by incorporating gradient information from previous epochs into the current gradient estimate. They alternate between model updates computed from small and large batches with a fixed pre-specified size. Two different batch sizes are concurrently used in a heterogeneous CPU+GPU setting in order to maximize utilization and reduce staleness for both resources~\cite{hetero-sgd:arxiv,hetero-sgd:ipdpsw-2021}. Unlike the variance reduction algorithms -- where model updates are sequential -- the CPU+GPU algorithm supports asynchronous model updates. Moreover, the two batch sizes are dynamically adapted -- not only increased -- based on the relative speed of the two processors. In this work, we extend the adaptive CPU+GPU algorithm to a multi-GPU setting where every GPU has its own batch size. The model merging strategy is more intricate and is performed synchronously after a variable number of epochs.

%%%%%%%%%%%%%%%%%%%%%%%%%%%%%%%%%%%%%%
\paragraph*{Training in Heterogeneous Settings}
Heterogeneity can manifest in multiple forms, including different types of processing units -- CPU and GPU -- on the same machine or clusters consisting of nodes with different characteristics. TensorFlow~\cite{tensorflow} and Omnivore~\cite{omnivore} support deep learning on heterogeneous CPU+GPU architectures. They schedule linear algebra primitives across CPU and GPU. The decision on where to perform a primitive depends on the estimated execution time for each device. Scheduling is heavily constrained by previous decisions because switching between CPU and GPU introduces time-consuming data transfers. Omnivore splits the training data into batches having size proportional to the speed of the device. The goal is to have perfectly synchronized execution with no delay across devices. The problem is that the actual speed at runtime can be quite different from the estimated one.

\medskip
\noindent
This issue is addressed by a heterogeneous CPU+GPU framework that leverages adaptive batch sizes and asynchronous model updates~\cite{hetero-sgd:arxiv,hetero-sgd:ipdpsw-2021}. In this framework, a collection of CPU threads designated as CPU workers perform asynchronous Hogwild SGD~\cite{hogwild} on a globally shared model, while the GPUs -- which perform mini-batch SGD on local model replicas -- are controlled by a separate set of CPU threads designated as GPU workers. The synchronization among all the workers is performed by a separate CPU coordinator thread that implements Hogbatch SGD~\cite{hogbatch} to update the global model. Empirical results show that including CPU workers in the SGD training workflow decreases the time-to-accuracy compared with the GPU- and CPU-only solutions. However, this holds for settings with one -- or at most two -- GPUs. When more than two GPUs are present, the ratio of training samples assigned to CPU workers becomes too small due to their significantly slower processing speed, essentially eliminating their contribution to SGD training.

\medskip
\noindent
The main difference in a distributed parameter server setting is that training data are statically partitioned to workers. Moving data between workers incurs expensive network traffic and is not viable. Instead, existing solutions use different learning rates across workers~\cite{hetero-ps} or train multiple models concurrently and choose the one that provides the highest accuracy~\cite{cerebro}. CROSSBOW~\cite{crossbow} considers heterogeneity in the context of multiple GPUs by scheduling a different number of processing streams -- learners -- on every GPU. However, model merging across GPUs does not consider the number of learners. Moreover, the reported experimental results in~\cite{crossbow} do not vary the number of learners across GPUs. The strategy proposed in this work varies the batch size for every GPU and merges the local models following a principled approach.

%%%%%%%%%%%%%%%%%%%%%%%%%%%%%%%%%%%%%%%%%%%%%%%%%%%%%%%%%%%%%%%%%
%\input{adaptive-elastic-sgd}
\section{ADAPTIVE ELASTIC SGD ON HETEROGENEOUS MULTI-GPUS}\label{sec:adap-elastic-sgd}

In this section, we present Adaptive SGD---a novel adaptive elastic model averaging SGD algorithm for heterogeneous multi-GPUs. We begin with a discussion of the high-level ideas outlining the structure of the algorithm and follow with the technical details. While we target a single-server setting, the proposed algorithm can be also applied in a distributed environment as long as training batches are dynamically scheduled across computing nodes---instead of being statically partitioned before processing.

%%%%%%%%%%%%%%%%%%%%%%%%%%%%%%%%%%%%%
\begin{figure}[htbp]
\centering
\includegraphics[width=\textwidth]{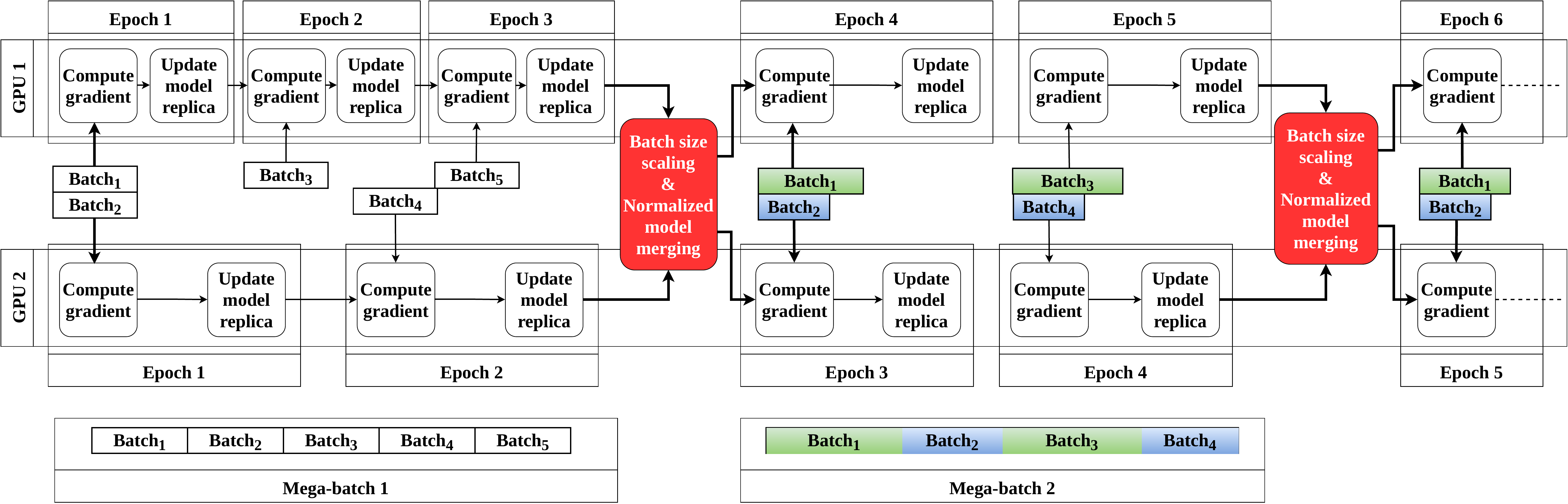}
\caption{Adaptive elastic SGD on heterogeneous multi-GPUs.}
\label{fig:adaptive-elastic-sgd}
\end{figure}
%%%%%%%%%%%%%%%%%%%%%%%%%%%%%%%%%%%%%

%%%%%%%%%%%%%%%%%%%%%%%%%%%%%%%%%%%%%%
\subsection{High-level Idea}\label{ssec:high-level}

Elastic model averaging (Figure~\ref{fig:elastic-model-avg}) imposes a strict requirement that every GPU has to process the same number of batches with the same size between two model averaging stages. This guarantees that every model replica is updated the same number of times with gradients approximated from samples with the same cardinality. Thus, averaging is straightforward. In a heterogeneous multi-GPU environment, this requirement exacerbates the straggler problem by increasing the waiting time of the faster GPUs and reducing their utilization. Ultimately, this increases the time-to-accuracy.

\medskip
\noindent
We address the limitation of elastic model averaging with continuous monitoring of the GPU execution, dynamic scheduling, and adaptive batch size scaling. Since monitoring is performed only at the synchronization points, its overhead is negligible because it overlaps with model transfer and merging. Instead of statically assigning batches to GPUs for every model averaging stage, in dynamic scheduling batches are dispatched one-by-one based on GPU availability. Whenever a GPU finishes an epoch, it is assigned the next batch. Similar to elastic model averaging, this process is controlled by fixing the number of training samples processed between model merging stages---we call these samples a mega-batch. As depicted in Figure~\ref{fig:adaptive-elastic-sgd}, dynamic scheduling can lead to a different number of model updates across GPUs---3 updates for GPU 1 and 2 updates for GPU 2. This creates an impediment for model averaging because the replicas are not on the same time horizon. In our example, the replica on GPU 1 is one step ahead of the replica on GPU 2. In order to mitigate the stale replica problem, we design an adaptive batch size scaling mechanism that is performed during model merging. The goal of batch size scaling is to assign larger batches to the faster GPUs and smaller batches to the slower ones, such that they perform the same number of model updates---as shown for the second mega-batch in Figure~\ref{fig:adaptive-elastic-sgd}. While this brings the replicas on the same time horizon, it also requires significant changes to the SGD algorithm. First, the batch size corresponding to every GPU has to be adequately determined. The dynamic modification of the batch size also requires an update of the corresponding learning rate. Thus, both of these hyperparameters become instances associated with every GPU, rather than being global for the SGD algorithm. This leads to the second change. Model merging has to carefully consider the differences across replicas, including the hyperparameters -- batch size and learning rate -- as well as the magnitude of the model parameters. We discuss the technical details of the resulting Adaptive SGD algorithm in the following sections.

%%%%%%%%%%%%%%%%%%%%%%%%%%%%%%%%%%%%%%%%%%%%%%%%%%%%%%
\begin{algorithm}[htbp]
	\caption{Batch Size Scaling}\label{alg:batch-size-scale}
	\begin{algorithmic}[1]

		\Statex {\bf Input:} current batch size $b_{i}$ and learning rate $\textit{lr}_{i}$ for every GPU $i$
		\Statex \hspace{.2cm} number of model replica updates $u_{i}$ for every GPU $i$ since the last model merging
		\Statex \hspace{.2cm} minimum $b_{\textit{min}}$ and maximum $b_{\textit{max}}$ allowable batch size
		\Statex \hspace{.2cm} batch size scaling parameter $\beta$
		
		\Statex {\bf Output:} updated batch size $b_{i}$ and learning rate $\textit{lr}_{i}$ for every GPU $i$

		\State Average number of model updates per GPU: $\tilde{\mu} \leftarrow \left(\sum_{i}{u_{i}}\right) / |\textit{GPU}|$

		\Statex $\rhd$ \textbf{Scale batch size and learning rate for the GPUs deviating from the average}
		\ForAll{$\textit{GPU}_{i}$}
	    \State \textbf{if} $u_i > \tilde{\mu}$ \textbf{and} $b_i + \beta \cdot \left( u_i - \tilde{\mu} \right) \leq b_{\textit{max}}$ \textbf{then}
				\Statex \hspace{1cm} $\rhd$ \textbf{Increase batch size and learning rate for faster GPUs}
        \State \hspace{1cm} $\textit{lr}_i \leftarrow \textit{lr}_i \cdot (b_i + \beta \cdot \left( u_i - \tilde{\mu} \right)) / b_i$
        \State \hspace{1cm} $b_i \leftarrow b_i + \beta \cdot \left( u_i - \tilde{\mu} \right)$
	    \State \textbf{else if} $u_i < \tilde{\mu}$ \textbf{and} $b_i - \beta \cdot \left( \tilde{\mu} - u_i \right) \geq b_{\textit{min}}$ \textbf{then}
				\Statex \hspace{1cm} $\rhd$ \textbf{Decrease batch size and learning rate for slower GPUs}
        \State \hspace{1cm} $\textit{lr}_i \leftarrow \textit{lr}_i \cdot (b_i - \beta \cdot \left( \tilde{\mu} - u_i \right)) / b_i$
        \State \hspace{1cm} $b_i \leftarrow b_i - \beta \cdot \left( \tilde{\mu} - u_i \right)$
      \State \textbf{end if}
		\EndFor

	\end{algorithmic}
\end{algorithm}
%%%%%%%%%%%%%%%%%%%%%%%%%%%%%%%%%%%%%%%%%%%%%%%%%%%%%%

%%%%%%%%%%%%%%%%%%%%%%%%%%%%%%%%%%%%%%
\subsection{Batch Size Scaling}\label{ssec:adap-bsize}

The batch size scaling procedure is presented in Algorithm~\ref{alg:batch-size-scale}. The goal is to arrive at a steady state in which all the GPUs perform the same number of replica updates. By default, the algorithm is executed after every mega-batch. However, if stability is achieved or the system enters an oscillatory state, the frequency at which scaling is performed can be increased. The algorithm targets to compute an updated batch size $b_i$ and learning rate $\textit{lr}_i$ for every GPU $i$ based on their corresponding number of model updates $u_i$. Since the update to the learning rate is completely determined by the linear scaling rule introduced in~\cite{facebook:large-batches}, the main challenge is how to update the batch size. After experimenting with several functions, we settle for linear update with a parameter $\beta$ determined empirically. The batch size $b_i$ is increased/decreased by a factor proportional to the deviation of the number of updates $u_i$ from the average number of updates $\tilde{\mu}$ across all GPUs.

\medskip
\noindent
In order to guarantee a minimum degree of GPU utilization, the minimum batch size is bounded by a threshold $b_{\textit{min}}$. A symmetric bound $b_{\textit{max}}$ controls the maximum size of a batch that fits in the GPU memory. $b_{\textit{min}}$ and $b_{\textit{max}}$ are derived from the mega-batch size such that the discrepancy between the number of model updates performed across GPUs is restricted. Otherwise, some GPUs dominate the SGD algorithm, rendering the others ineffective---they can be removed without impacting the SGD time-to-accuracy. Moreover, $b_{\textit{min}}$ and $b_{\textit{max}}$ allow for analytical characterization of batch size scaling. Assuming an equal number of model updates across GPUs, the convergence behavior of SGD with batch size scaling is within the range of elastic model averaging with a batch size of $b_{\textit{min}}$ and $b_{\textit{max}}$, respectively. When the number of model updates varies, these thresholds impose bounds on replica staleness, which allows for the application of convergence results from stale synchronous SGD~\cite{stale-synch-ps,asynch-elastic-sgd}.

%%%%%%%%%%%%%%%%%%%%%%%%%%%%%%%%%%%%%%%%%%%%%%%%%%%%%%
\begin{algorithm}[htbp]
	\caption{Normalized Model Merging}\label{alg:norm-model-merge}
	\begin{algorithmic}[1]

		\Statex {\bf Input:} current global model $\overline{w}$, previous global model $\overline{w_{p}}$, model replica $\overline{w}_{i}$ for every GPU $i$
		\Statex \hspace{.2cm} batch size $b_{i}$ and learning rate $\textit{lr}_{i}$ for every GPU $i$
		\Statex \hspace{.2cm} number of model replica updates $u_{i}$ for every GPU $i$ since the last model merging
		\Statex \hspace{.2cm} perturbation threshold $\textit{pert}_{\textit{thr}}$, perturbation factor $\delta$
		\Statex \hspace{.2cm} momentum hyperparameter $\gamma$
		
		\Statex {\bf Output:} updated global model $\overline{w}$

		\Statex $\rhd$ \textbf{Compute the normalization weights}
		\State Let $\alpha_{i}$ be the weight corresponding to every GPU $i$

		\If{$u_{i} = u_{j}, \forall i, j \in \textit{GPU}$} \hspace{.3cm} $\rhd$ \textbf{Normalize weights based on batch size}
			\State $\alpha_{i} \leftarrow b_{i} / \left( \sum_{i}{b_{i}} \right), \forall i \in \textit{GPU}$
		\Else \hspace{.3cm} $\rhd$ \textbf{Normalize weights based on number of model updates}
			\State $\alpha_{i} \leftarrow u_{i} / \left( \sum_{i}{u_{i}} \right), \forall i \in \textit{GPU}$
		\EndIf
		
		\Statex $\rhd$ \textbf{Add perturbation to min/max normalization weights when all replicas are regularized}
		\If{$||\overline{w}_{i}||_{2} / |\overline{w}| < \textit{pert}_{\textit{thr}}, \forall i \in \textit{GPU}$}
			\State $r \leftarrow \operatorname{argmax}_{i} \left\{ {u_{i}} \right\}$, $s \leftarrow \operatorname{argmin}_{i} \left\{ {u_{i}} \right\}$, $\forall i \in \textit{GPU}$
			\State $\alpha_{r} \leftarrow (1 + \delta) \cdot \alpha_{r}$, $\alpha_{s} \leftarrow (1 - \delta) \cdot \alpha_{s}$
		\EndIf

		\Statex $\rhd$ \textbf{Compute and update the global model}
		\State $\overline{w'} \leftarrow \sum_{i}{\alpha_{i} \cdot \overline{w}_{i}} + \gamma \cdot \left( \overline{w} - \overline{w_{p}} \right)$
		\State $\overline{w_{p}} \leftarrow \overline{w}$, $\overline{w} \leftarrow \overline{w'}$

	\end{algorithmic}
\end{algorithm}
%%%%%%%%%%%%%%%%%%%%%%%%%%%%%%%%%%%%%%%%%%%%%%%%%%%%%%

%%%%%%%%%%%%%%%%%%%%%%%%%%%%%%%%%%%%%%
\subsection{Normalized Model Merging}\label{ssec:weighted-merge}

While batch size scaling aims to arrive at the same number of model replica updates across all the GPUs, this cannot be achieved instantaneously---or at all. Thus, model merging has to consider both cases with and without identical number of updates when deriving the global model from the local replicas. Moreover, merging has to account for the difference in batch size. Algorithm~\ref{alg:norm-model-merge} is a general model merging procedure that handles these cases methodically. The underlying principle is to prioritize the replicas updated more frequently and -- secondarily -- with gradients derived from larger batch sizes. This leads to weighted average model merging, where the weights are normalized either based on the batch size -- when the number of model updates are identical -- or the number of updates itself. In the first case, larger batch sizes generate more accurate gradients, thus, more accurate models. For the second case, the choice to normalize exclusively based on the number of model updates requires some discussion. This situation is more likely to appear for the first mega-batches, before batch size scaling guides the Adaptive SGD algorithm to a steady state. Since the model is farther from the global minimum, a wider exploration is beneficial to identify the appropriate batch size. This is similar to the warmup approach~\cite{facebook:large-batches}. An alternative for later stages is to normalize based on the product between the number of updates and the batch size. While this increases the weight of replicas updated with larger batches, empirical results do not show a discernible improvement. Therefore, we adopt normalization based exclusively on the number of model updates in Algorithm~\ref{alg:norm-model-merge}.

\medskip
\noindent
In order to increase the importance of the most updated replica, Algorithm~\ref{alg:norm-model-merge} adds perturbations to the normalized weights. This is done by increasing the weight corresponding to the most updated replica and decreasing the weight for the replica with the fewest updates accordingly. The degree of increase/decrease is controlled by the perturbation factor $\delta$, which is a parameter defined by the user and set by default to $0.1$. It is important to notice that this weight modification can result in denormalization---the sum of the weights $\alpha$ is not equal to $1$ anymore. In order to restrict the eventual impact of denormalization, we apply weight perturbation only when all the model replicas are well regularized. We quantify model regularization by the L2-norm per model parameter measure, i.e., L2-norm divided by model dimensionality. Since large values of the L2-norm correspond to unregularized models -- skewed along one or more dimensions -- weight perturbation is applied only when the L2-norm per model parameter is below a threshold $\textit{pert}_{\textit{thr}}$ for all the replicas---$\textit{pert}_{\textit{thr}}$ is set by default to $0.1$ and can be tuned by the user. This ensures that no skewed parameters are amplified by the denormalized weights.

\medskip
\noindent
The final step in Algorithm~\ref{alg:norm-model-merge} is to update the global model. This process consists in computing the weighted average of the local replicas using the normalized -- and perturbed -- weights. Rather than directly assigning this value to the global model, we follow the SGD momentum update rule~\cite{sgd:survey}, in which the current and previous global model are combined with the merged replicas. Momentum preserves a stable convergence trajectory while avoiding reactive oscillations. The scale of the past models is controlled by the momentum hyperparameter $\gamma$---set to $0.9$ according to the literature~\cite{sgd:survey}.

%%%%%%%%%%%%%%%%%%%%%%%%%%%%%%%%%%%%%%%%%%%%%%%%%%%%%%%%%%%%%%%%%
%\input{system}
\section{IMPLEMENTATION CONSIDERATIONS}\label{sec:implementation}

The Adaptive SGD algorithm for heterogeneous multi-GPUs is implemented in the HeteroGPU framework for sparse deep learning. The open-source code is available on Github~\cite{hetero-gpu:github}. In this section, we present the HeteroGPU architecture and workflow. Then, we discuss a series of optimizations to enhance the performance of the GPU code. These include CUDA kernel fusion, multi-stream all-reduce model merging functions, and scalable memory management.

%%%%%%%%%%%%%%%%%%%%%%%%%%%%%%%%%%%%%
\begin{figure}[htbp]
\centering
\includegraphics[width=\textwidth]{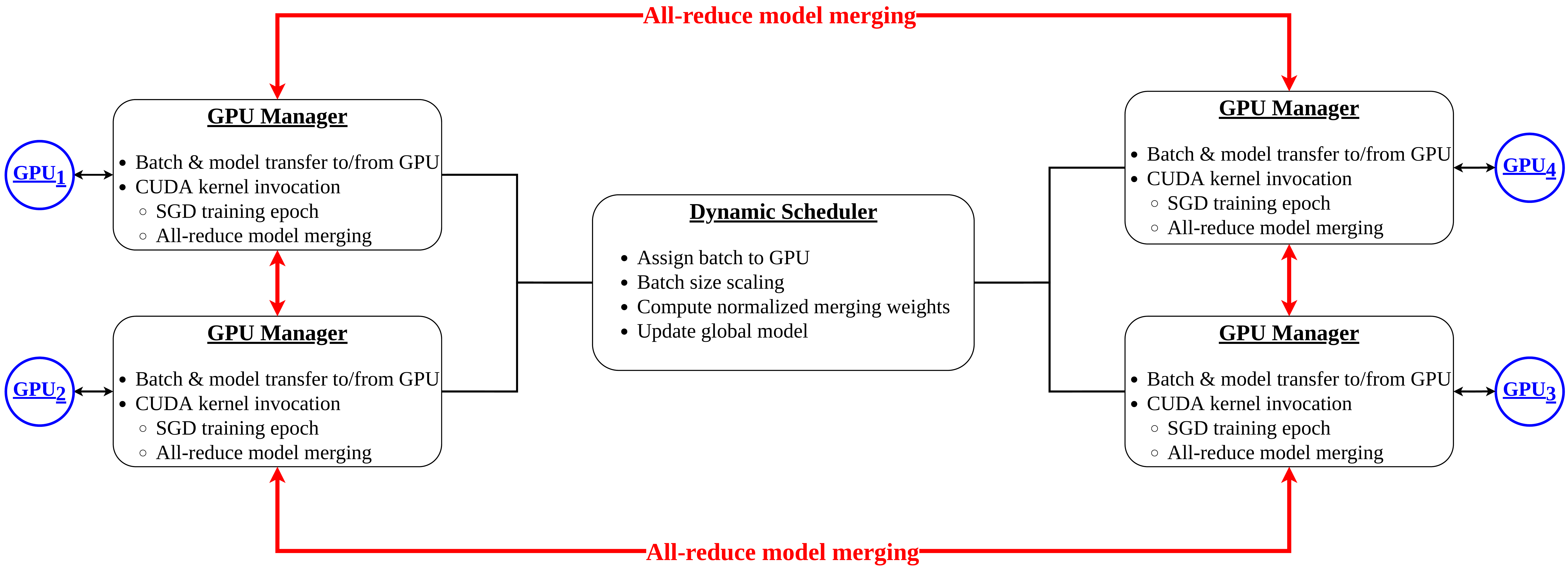}
\caption{System implementation architecture and workflow.}
\label{fig:system-arch-flow}
\end{figure}
%%%%%%%%%%%%%%%%%%%%%%%%%%%%%%%%%%%%%

%%%%%%%%%%%%%%%%%%%%%%%%%%%%%%%%%%%%%%
%\subsection{Architecture and Workflow}\label{ssec:architecture}
\paragraph*{Architecture and Workflow}
The architecture of HeteroGPU together with the main tasks performed by every component are depicted in Figure~\ref{fig:system-arch-flow}. HeteroGPU consists of multiple asynchronous GPU managers corresponding to every GPU in the system -- four in this example -- and a central dynamic scheduler. They are implemented as stand-alone asynchronous threads that communicate through event messages.

\medskip
\noindent
The GPU manager is in charge of the resources and execution of its assigned GPU. It coordinates the data transfers between CPU and GPU, and invokes the CUDA kernels. The execution of an epoch of the SGD algorithm requires moving the training batch and the model replica -- only at the beginning of a mega-batch -- to the GPU and then invoking linear algebra kernels for gradient computation and replica update. Although the GPU manager provides thread isolation for these operations, there is a significant degree of interference in the CUDA environment scheduler -- which is shared across GPUs -- due to multiple GPU managers launching CUDA kernels simultaneously. The interference manifests by a large kernel startup overhead---which increases with the number of GPUs. HeteroGPU addresses this issue by kernel fusion in event-based asynchronous execution streams. Specifically, small kernels that perform primitive operations, e.g., element-wise linear algebra, are grouped into a large kernel. The GPU manager invokes this kernel in an independent stream and monitors its completion using events that bypass the global CUDA environment. The intermediate output of kernel invocations is kept in the GPU memory in order to reduce data movement. Nonetheless, advanced memory management strategies that work at hidden layer granularity~\cite{memory-mgmt:ipdps-2019} can also be supported. The other major task performed by the GPU manager is model merging at mega-batch boundaries. This is realized with an all-reduce procedure that involves direct data transfers among GPUs.

\medskip
\noindent
The most common task of the dynamic scheduler is to assign data batches of different size to the GPU managers. The size of the batches is determined in the batch size scaling procedure---performed at the end of processing a mega-batch. The computation of the normalized model merging weights is also performed at this stage. As shown in Algorithm~\ref{alg:batch-size-scale} and~\ref{alg:norm-model-merge}, these require the number of model replica updates executed by every GPU manager---which are recorded by the scheduler when batches are dispatched. The overhead of these tasks is reduced, making the scheduler a relatively low utilized component. While model merging is performed exclusively by the GPU managers -- with the normalized weights computed by the scheduler -- model update requires the combination of the merged model with the current and previous global model. This operation can be executed either on every GPU or by the scheduler. We opt for the second alternative because it requires fewer data transfers between CPU and GPU. Moreover, since the operations are element-wise, the computation time is low.

%%%%%%%%%%%%%%%%%%%%%%%%%%%%%%%%%%%%%%
%\subsection{All-reduce Model Merging}\label{ssec:all-reduce}
\paragraph*{All-reduce Model Merging}
Model merging is implemented as an all-reduce operation in the HeteroGPU framework. This diminishes the load on the dynamic scheduler, avoiding eventual bottlenecks. In all-reduce model merging, the aggregate -- in this case weighted average -- is computed through multi-round data partitioning such that all the GPUs end up with the overall aggregated model. While the NCCL~\cite{nvidia-nccl} multi-GPU communication library provides all-reduce methods, they lack support for multi-streams -- which precludes the overlap between model transfer and reduction computation -- or are optimized for a multi-server setting---which is not compatible with our single-server scope. Therefore, we implement specialized tree- and ring-based multi-stream all-reduce aggregation functions. The local replica models are split into a fixed number of partitions, which are allocated to a separate GPU processing stream. The optimal number of partitions -- and GPU streams -- is empirically determined to be equal with the number of GPUs in the system. Every stream performs the all-reduce aggregation starting from a different GPU. This results in complete overlap between data transfer and computation. While the NCCL tree-based implementation is more efficient on a single stream, the multi-stream ring-based all-reduce function is faster than its corresponding tree-based equivalent. Thus, this is the method used throughout the experiments.

%%%%%%%%%%%%%%%%%%%%%%%%%%%%%%%%%%%%%%
%\subsection{Scalable Memory Management}\label{ssec:mem-management}
\paragraph*{Scalable Memory Management}
Two memory management issues in HeteroGPU require discussion. The first is effective memory allocation. We maintain a memory pool for all the memory allocation operations. The training dataset is loaded in a sparse representation in the CPU memory during initialization. We also allocate memory for the model instances and gradients on GPUs at initialization and use it during the entire execution. In particular, the memory for the partial results in every hidden layer is reused during testing and validation. In order to handle adaptive batch sizes, we provision memory for the largest batch size. Nonetheless, only the necessary size is assigned to an operation at runtime. Therefore, there is no memory allocation overhead in HeteroGPU since everything is pre-allocated. The second issue is related to the memory usage in the multi-stream all-reduce methods. In a naive implementation, there is a model replica instance for every GPU stream. This is prohibitive due to the large memory footprint. Since the merged model is obtained as a weighted average of the replicas, the only quantity that has to be maintained is the partial average across the already included GPUs. This requires at most an additional model instance. In fact, it can even be done directly on the model replicas at the cost of overwriting the current content. With these optimizations, the overhead from memory usage and allocation is minimized.

%%%%%%%%%%%%%%%%%%%%%%%%%%%%%%%%%%%%%%%%%%%%%%%%%%%%%%%%%%%%%%%%%
%\input{experiments}
\section{EXPERIMENTAL EVALUATION}\label{sec:experiments}

We pursue three objectives in the experimental evaluation. First, we assess the overall performance of the proposed Adaptive SGD by comparing it with four well-known baseline algorithms. Second, we identify optimal values for the Adaptive SGD's parameters included in Algorithm~\ref{alg:batch-size-scale} and~\ref{alg:norm-model-merge}. Finally, we check if the specific heterogeneous multi-GPU characteristics of Adaptive SGD have a tangible effect on training. To this end, the specific questions we ask in the experiments are the following:
\begin{itemize}[noitemsep,nolistsep]
\item How does Adaptive SGD compare with Elastic SGD, TensorFlow, CROSSBOW, and SLIDE in terms of time-to-accuracy and statistical efficiency?
\item How scalable is Adaptive SGD with the number of GPUs?
\item How often has model merging to be performed for optimal Adaptive SGD performance?
\item What are the optimal values for the batch size scaling parameters in Algorithm~\ref{alg:batch-size-scale} and the perturbation parameters in Algorithm~\ref{alg:norm-model-merge}?
\item Do batch size scaling and perturbed model merging get invoked in practice and how often?
\end{itemize}

%%%%%%%%%%%%%%%%%%%%%%%%%%%%%%%%%%%%%
\subsection{Setup}\label{ssec:experiments:setup}

%%%%%%%%%%%%%%%%%%%%%%%%%%%%%%%%%%%%%
\paragraph*{Baseline Methods}
We include four alternative methods in the experimental evaluation. Three of them are multi-GPU SGD algorithms that do not consider heterogeneity, while the fourth is the SLIDE~\cite{slide-code} system, which provides a CPU-optimized SGD algorithm for sparse data. The three multi-GPU alternatives are TensorFlow~\cite{tensorflow-web} -- which implements gradient aggregation (Figure~\ref{fig:gradient-agg}) -- elastic model averaging (Figure~\ref{fig:elastic-model-avg}), and the CROSSBOW~\cite{crossbow} synchronous model averaging. We follow the experimental testbed for XML classification used in the SLIDE system. In addition to the SLIDE implementation, this testbed includes TensorFlow single-GPU code, which we extend to multi-GPUs both with the mirrored and central storage strategy. Since the mirrored strategy proves superior, we include only these TensorFlow results in the paper. We implement the remaining three methods -- CROSSBOW, elastic model averaging (Elastic SGD), and the proposed Adaptive SGD -- in the HeteroGPU framework~\cite{hetero-gpu:github}, which is a C++ prototype. We do not use the original CROSSBOW implementation because it lacks support for sparse data. The GPU code is written in CUDA 11.2, with primitives from the NVIDIA cuSPARSE library for the sparse linear algebra operations. The training data are stored in the sparse libSVM format.

%%%%%%%%%%%%%%%%%%%%%%%%%%%%%%%%%%%%%
\begin{table}[htbp]
    \centering
		\resizebox{\textwidth}{!}{
    \begin{tabular}{l|rrrrrr}
    \textbf{dataset} & \textbf{features} & \textbf{classes} & \textbf{training samples} & \textbf{testing samples} & \textbf{avg features per sample} & \textbf{avg classes per sample} \\
    \hline
    Amazon-670k & 135,909 & 670,091 & 490,449 & 153,025 & 76 & 5 \\
    Delicious-200k & 782,585 & 205,443 & 196,606 & 100,095 & 302 & 75 \\
    \end{tabular}
    }
    \caption{Experimental datasets for XML classification.}\label{tbl:dataset}
\end{table}
%%%%%%%%%%%%%%%%%%%%%%%%%%%%%%%%%%%%%

%%%%%%%%%%%%%%%%%%%%%%%%%%%%%%%%%%%%%
\paragraph*{Datasets and Model}
We use two standard datasets for XML classification in our experiments---Amazon-670k and Delicious-200k~\cite{Xclassification-dataset-repo}. Their characteristics are displayed in Table~\ref{tbl:dataset}. Both datasets have a very large number of classes and high-dimensional sparse input features, as shown by the average number of non-zero classes/features per sample. They are used to evaluate the performance of the SLIDE algorithm on a 3-layer Multi-Layer Perceptron (MLP) model having ReLU layer activation, softmax multi-class probability, and cross-entropy loss function. In order to maintain compatibility, we keep the same model configuration in our experiments. The best XML algorithms -- which are considerably more complex than the model used in these experiments -- achieve top-1 accuracy below 50\% on these datasets, while requiring hours to train and GB-size models~\cite{Xclassification-dataset-repo}. Thus, the presented results have to be interpreted in this context.

%%%%%%%%%%%%%%%%%%%%%%%%%%%%%%%%%%
%\textbf{Methodology.}
\paragraph*{Methodology}
We execute every algorithm for the same amount of time. We measure the top-1 accuracy -- corresponding to the class with the highest probability -- on the testing dataset after processing every mega-batch. This allows us to compute the time-to-accuracy, statistical efficiency, and hardware efficiency. All the algorithms are initialized with the same model, which gives the same initial loss. The initial values of the model weights are randomly drawn from a normal distribution with standard deviation equal to the number of units in every layer. The initial batch size -- set to $b_{\textit{max}}$ -- is chosen such that the GPU memory -- and utilization -- are maximized. $b_{\textit{min}}$ is set to a value 8 times smaller than $b_{\textit{max}}$, while the batch size scaling parameter $\beta$ to half of $b_{\textit{min}}$. The optimal learning rate for $b_{\textit{max}}$ is found by griding its range in powers of 10 and selecting the value that achieves the best accuracy across all the algorithms. The learning rates for the other batch sizes are determined based on the linear scaling rule~\cite{facebook:large-batches}. The same hyperparameters are used for all the algorithms, except for gradient aggregation in TensorFlow, where the batch size is decreased proportionally to the number of GPUs in order to compensate for the reduced number of model updates. The time to load the data and evaluate the accuracy are not included in timing measurements.

%%%%%%%%%%%%%%%%%%%%%%%%%%%%%%%%%%%%%
\paragraph*{Hardware}
We execute the experiments on a server with 4 NVIDIA Volta V100 GPUs, each with 16 GB of RAM. The server has a 16-core (32 threads) Intel 6226R Cascade Lake CPU with 192 GB of RAM and is running the Ubuntu 16.04.7 operating system. The number of GPUs used in an experimental trial is configured for every program separately.

%%%%%%%%%%%%%%%%%%%%%%%%%%%%%%%%%%%%%
\begin{figure}[htbp]
\centering
	\includegraphics[width=.32\textwidth]{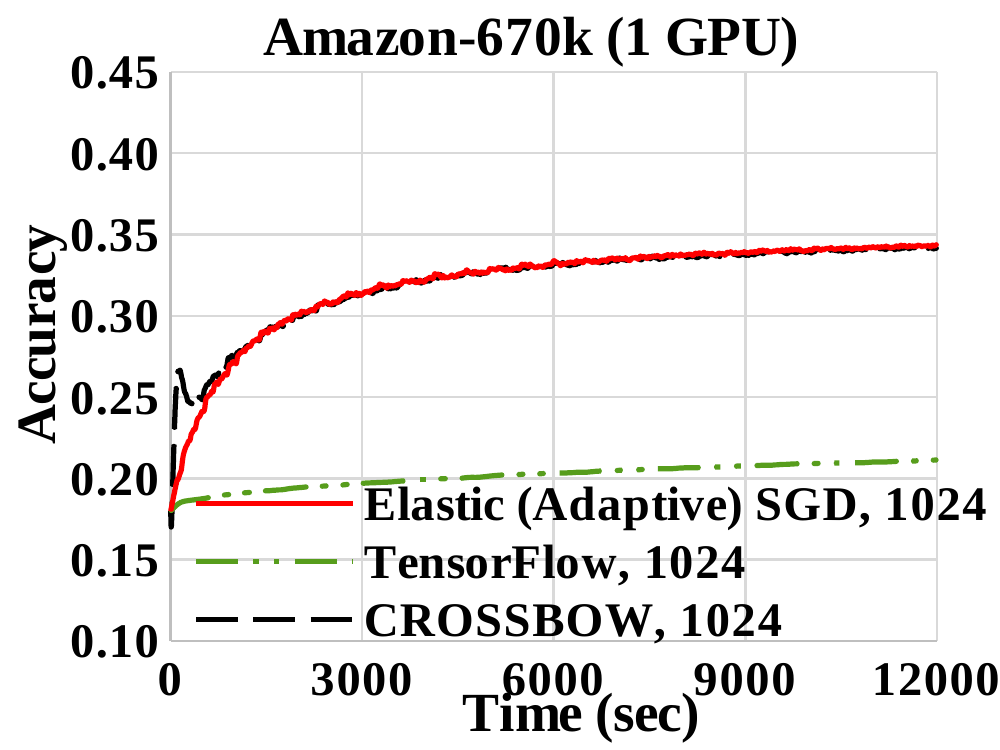}
	\includegraphics[width=.32\textwidth]{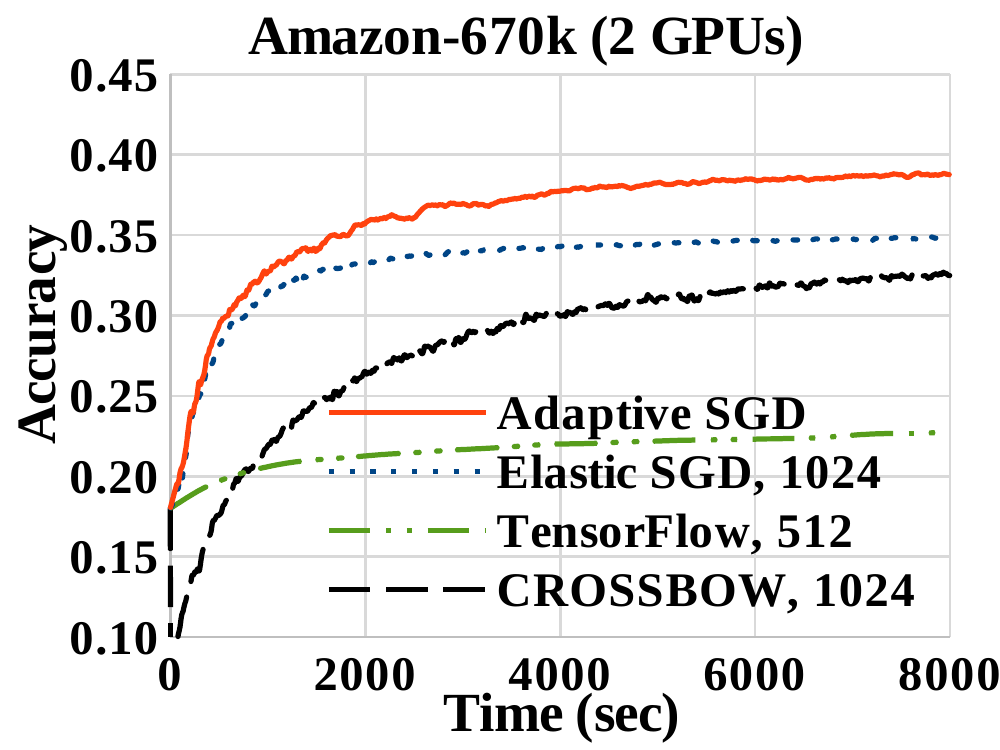}
	\includegraphics[width=.32\textwidth]{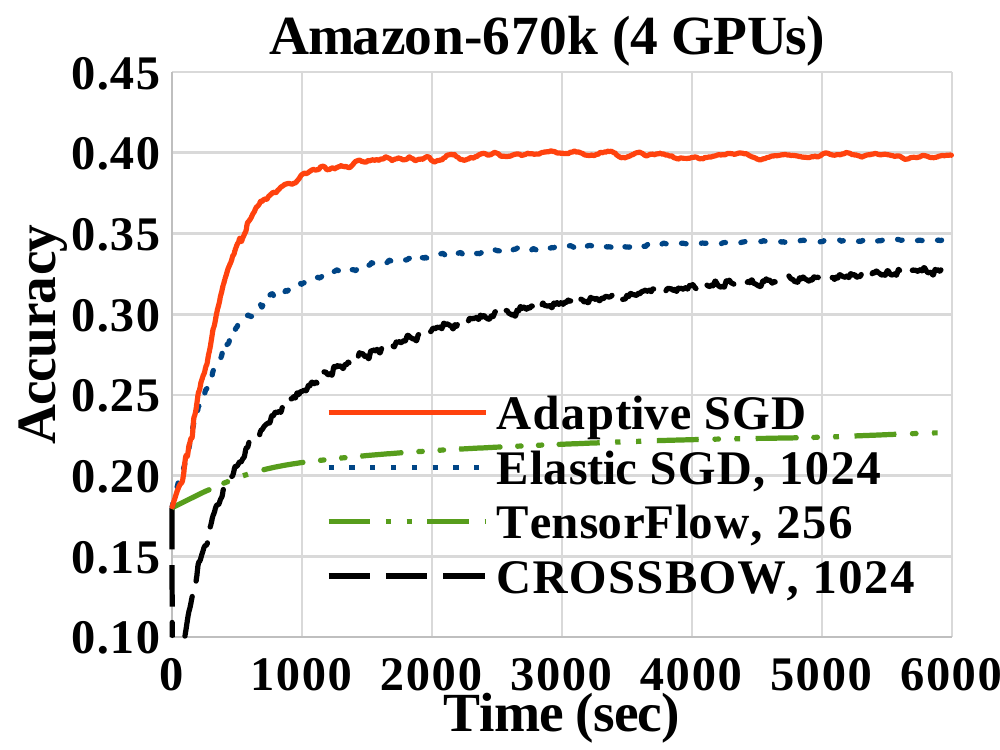}
	\\
	\includegraphics[width=.32\textwidth]{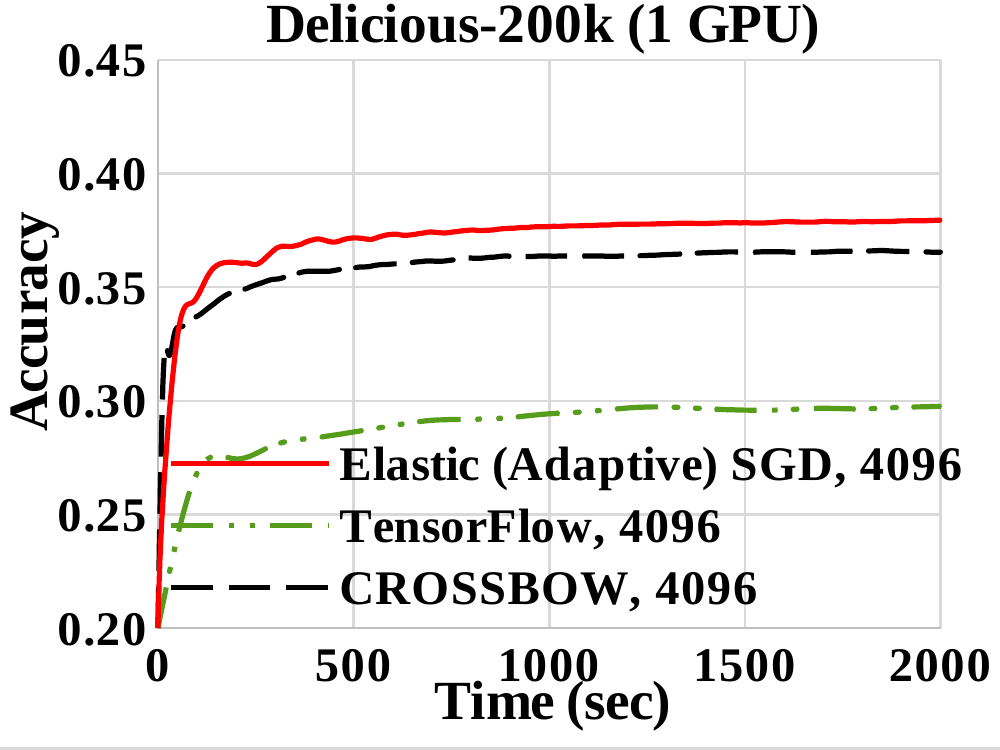}
	\includegraphics[width=.32\textwidth]{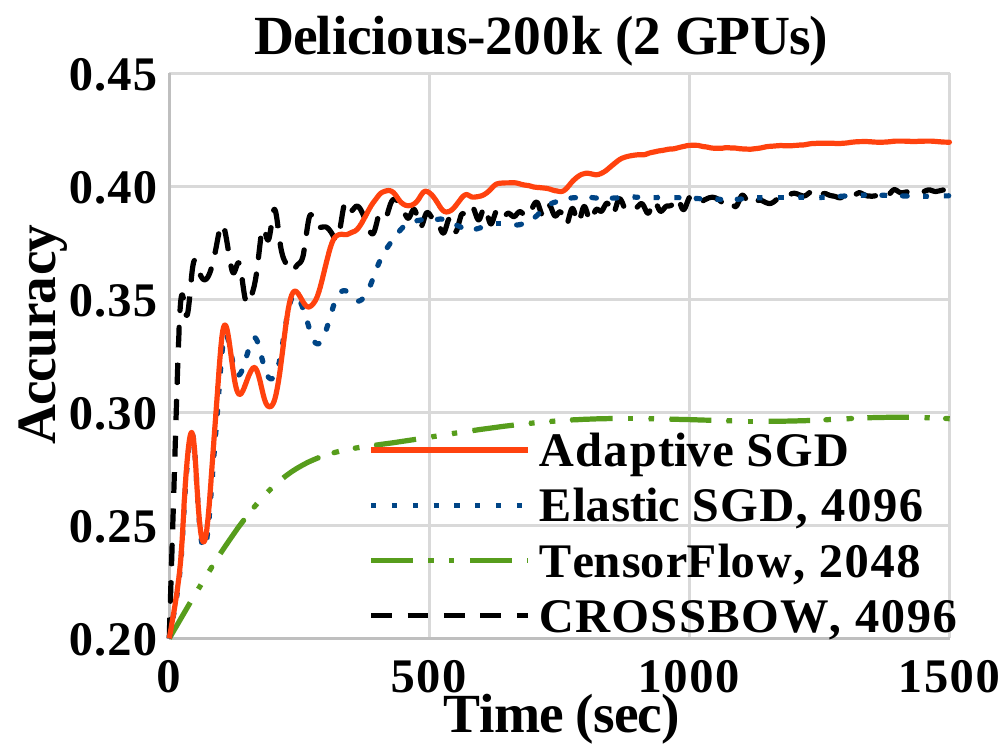}
	\includegraphics[width=.32\textwidth]{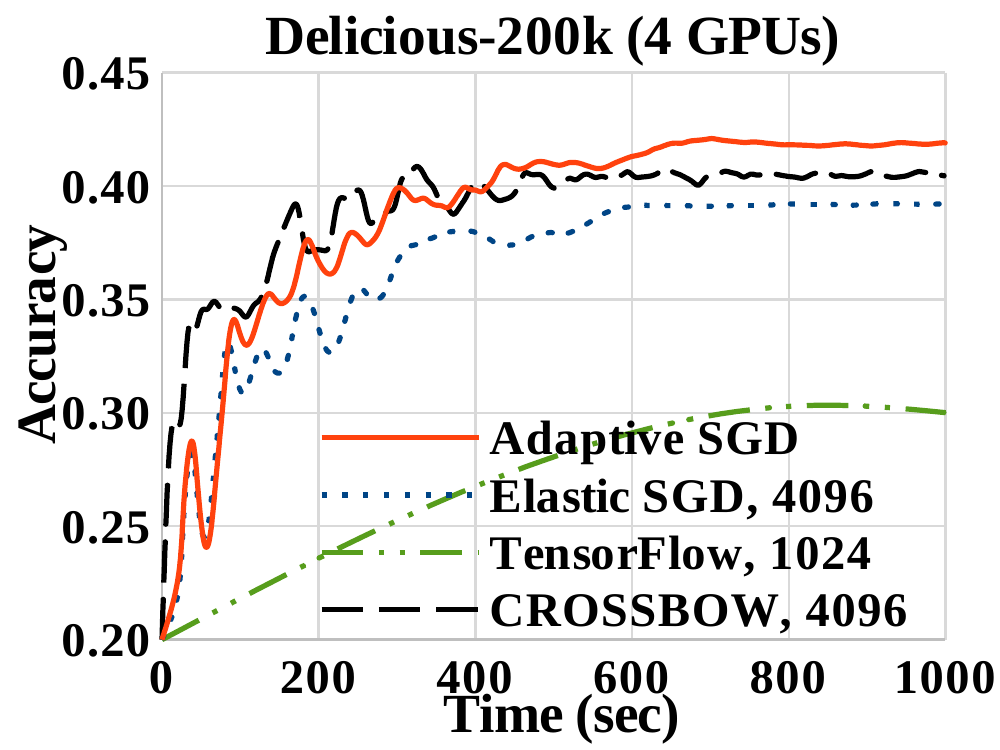}
\caption{Time-to-accuracy for a given number of GPUs.}
\label{fig:time-acc}
\end{figure}
%%%%%%%%%%%%%%%%%%%%%%%%%%%%%%%%%%%%%

%%%%%%%%%%%%%%%%%%%%%%%%%%%%%%%%%%%%%
\subsection{Results}\label{ssec:experiments:results}

We organize the results in benchmarks that compare the proposed Adaptive SGD algorithm with alternative solutions and micro-benchmarks that assess the impact of the algorithm's parameters on its behavior. While the former provide a complete picture on the applicability of Adaptive SGD, the later guide the user on how to set the different parameters of the algorithm.

%%%%%%%%%%%%%%%%%%%%%%%%%%%%%%%%%%%%%
\subsubsection{Overall Performance Benchmarks}

%%%%%%%%%%%%%%%%%%%%%%%%%%%%%%%%%%%%%
\paragraph*{Time-to-accuracy}
The time-to-accuracy measure is the most relevant metric to evaluate a training algorithm because it assesses the wall-clock time to achieve a certain level of accuracy. The shorter this time is, the less resources an algorithm uses. Figure~\ref{fig:time-acc} depicts time-to-accuracy for all the considered GPU methods. When the testing configuration has a single GPU, all the methods become mini-batch SGD. Since Elastic and Adaptive SGD are both implemented in HeteroGPU and use the same model update rule, they are identical---the reason they are depicted by a single curve. The model update in CROSSBOW includes the deviation of the local replica from the global model, thus, the different behavior. The TensorFlow curve is completely separate from the others because of the different implementation. There are several common trends across the two datasets. Adaptive SGD achieves the highest accuracy among all the methods in the shortest interval of time. This is the case for all the GPU configurations. The improvement over Elastic GPU proves the benefits of variable batch sizes and weighted model averaging. Without these, the delay from the slower GPUs increases the time to process a mega-batch. Moreover, the variation in batch size adds more variability to the local model replicas, which reflects in a more generalizable global model. Another trend is TensorFlow's considerably slower time-to-accuracy. There are two reasons for this. First, the execution of an SGD epoch and mirrored all-reduce aggregation are slower. Second, the global model is updated after every batch. In the case of Adaptive and Elastic SGD, the global model is updated only after a mega-batch having the size of 100 batches. CROSSBOW displays the most variability across the two datasets. It achieves much better accuracy on Delicious-200k than on Amazon-670k. This discrepancy is due to the sensitive global model update that can lead to divergent local replicas. There are two manifestations of the divergence. First, there is poor accuracy---the case for Amazon-670k. Second, there is instability---the case for Delicious-200k. While periodic replica recalibration can alleviate the divergence, the trigger and frequency of this operation are not well-defined. Nonetheless, when the accuracy values are relatively stable, Adaptive SGD achieves noticeably higher accuracy than CROSSBOW---especially for Amazon-670k.

%%%%%%%%%%%%%%%%%%%%%%%%%%%%%%%%%%%%%
\begin{figure}[htbp]
\centering
	\includegraphics[width=.32\textwidth]{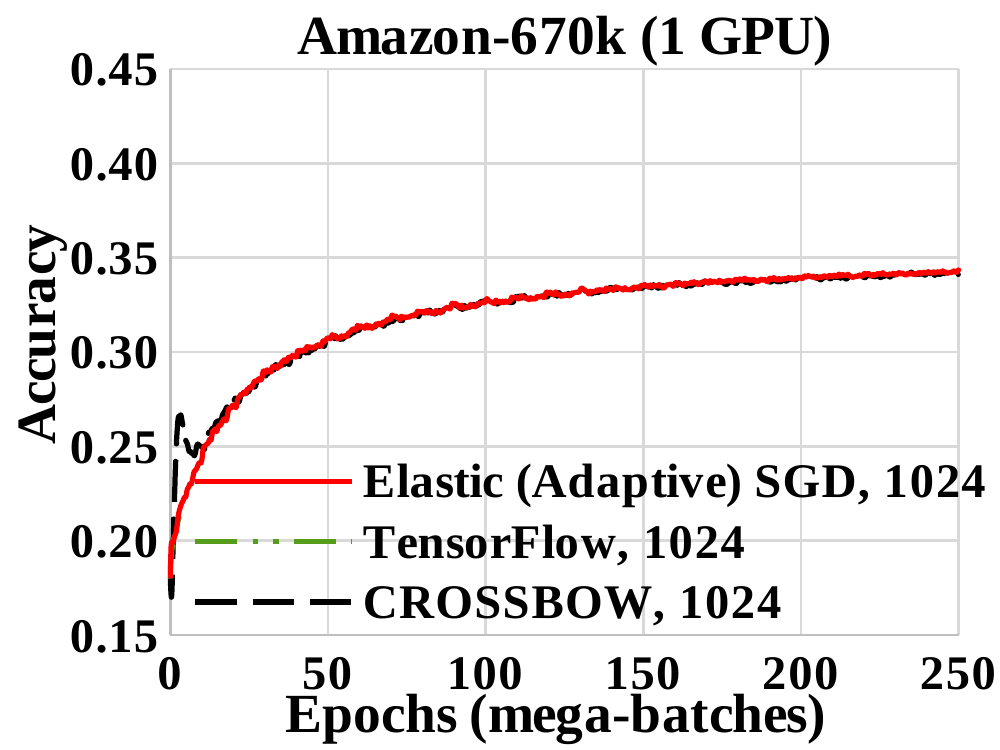}
	\includegraphics[width=.32\textwidth]{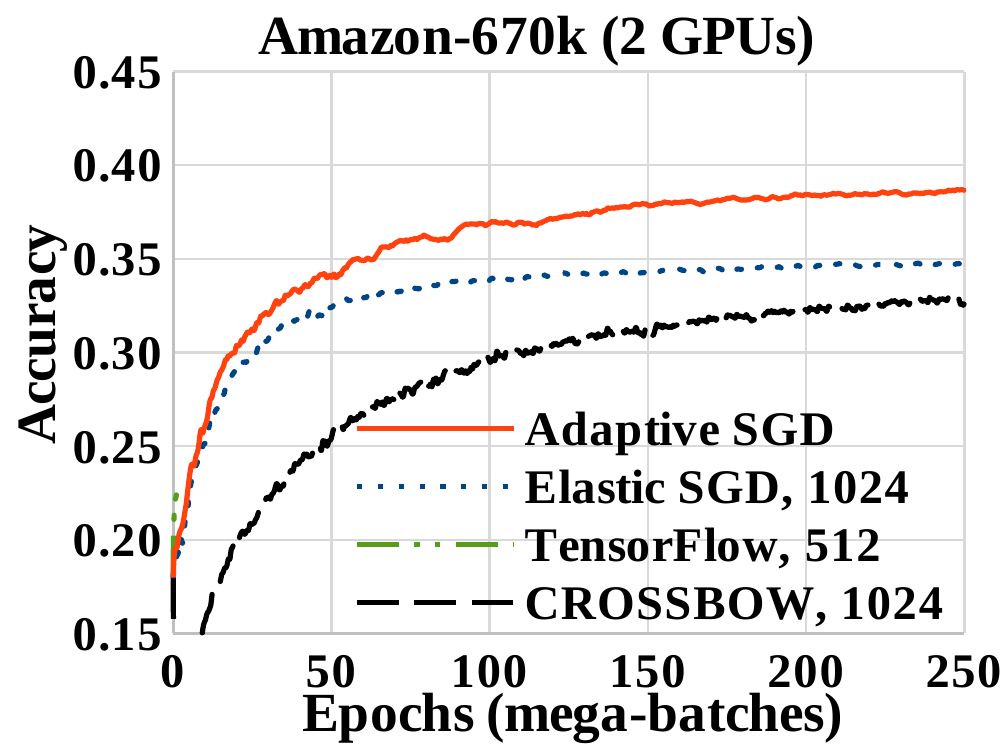}
	\includegraphics[width=.32\textwidth]{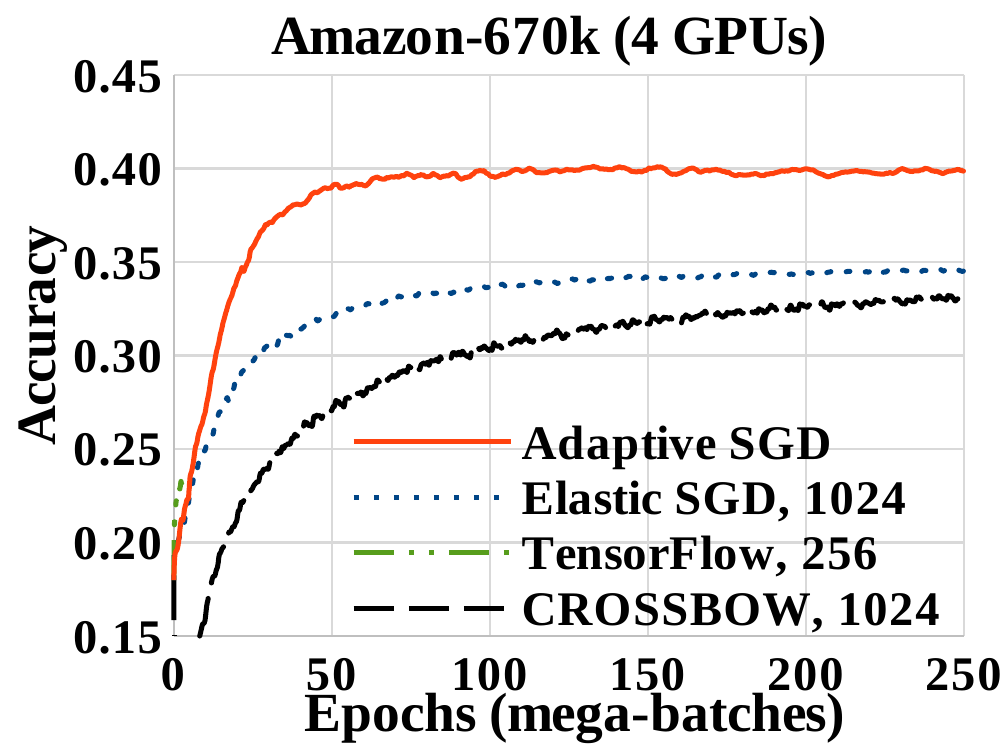}
	\\
	\includegraphics[width=.32\textwidth]{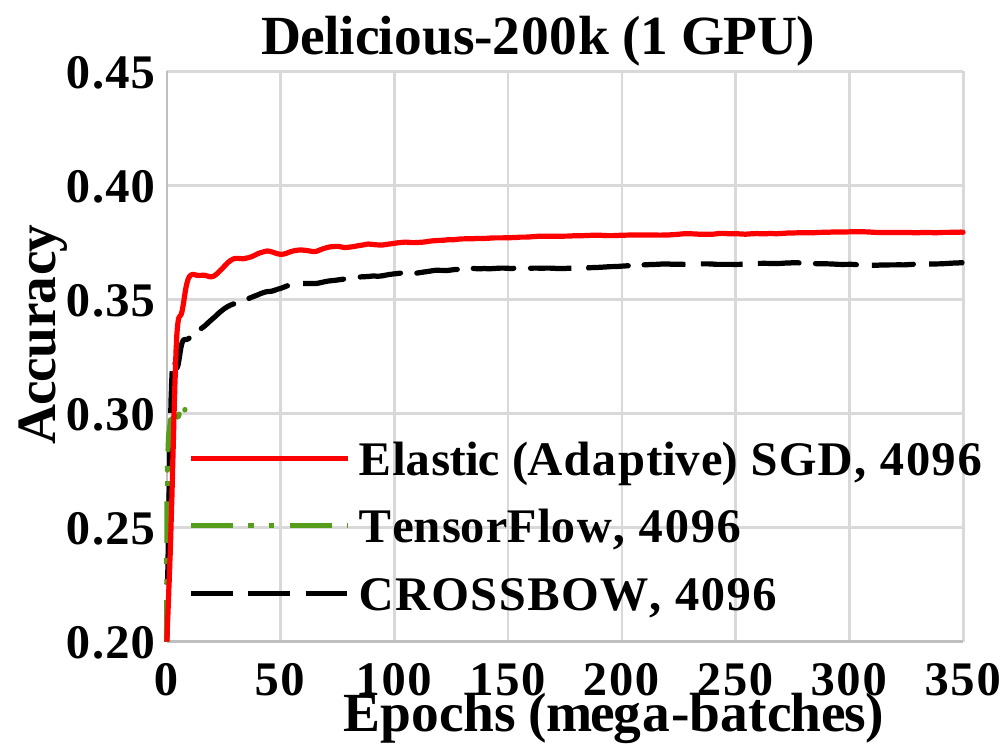}
	\includegraphics[width=.32\textwidth]{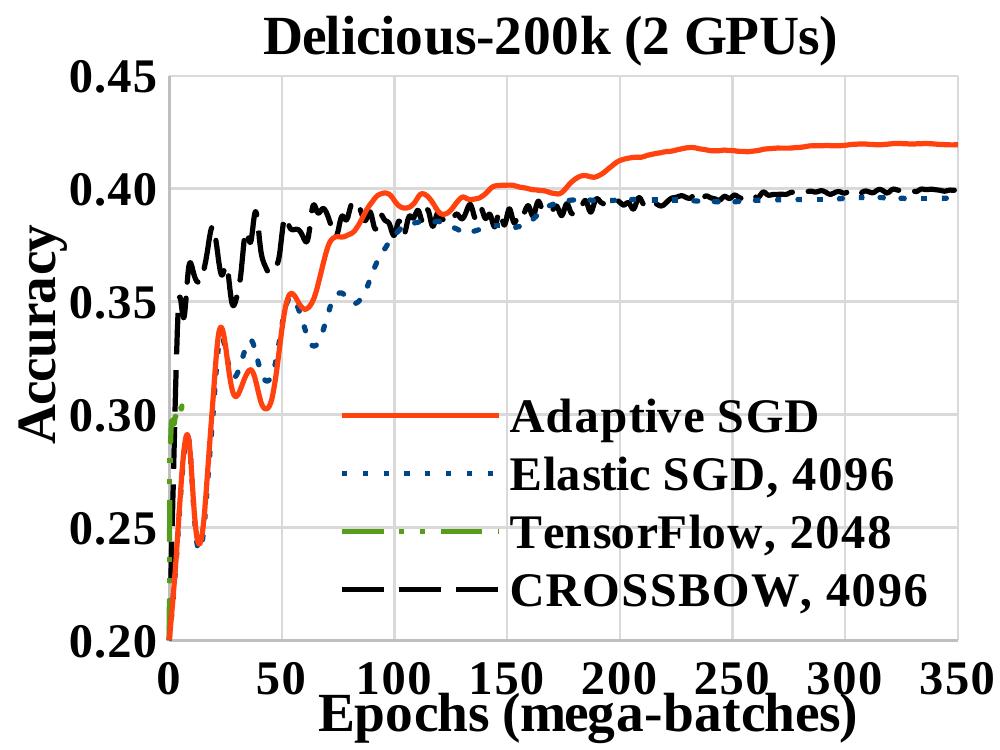}
	\includegraphics[width=.32\textwidth]{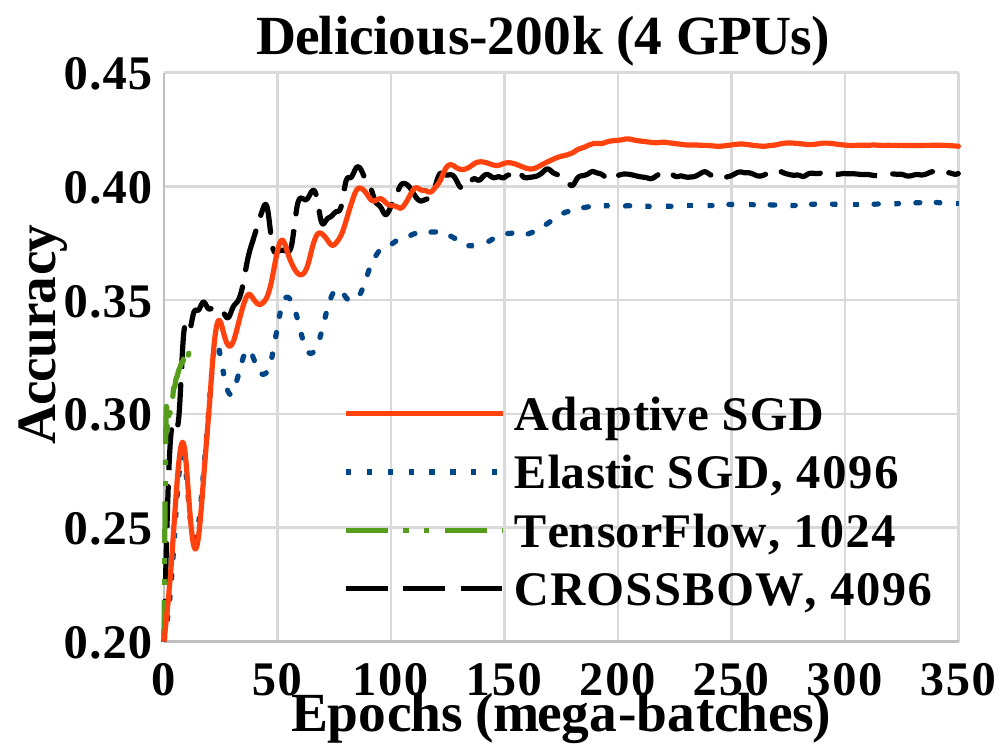}
\caption{Statistical efficiency for a given number of GPUs. An epoch corresponds to a mega-batch.}
\label{fig:stat-eff}
\end{figure}
%%%%%%%%%%%%%%%%%%%%%%%%%%%%%%%%%%%%%

%%%%%%%%%%%%%%%%%%%%%%%%%%%%%%%%%%%%%
\paragraph*{Statistical efficiency}
The plots for statistical efficiency in Figure~\ref{fig:stat-eff} -- which depict accuracy as a function of the number of mega-batches -- have a similar pattern as the curves for time-to-accuracy. The only significant difference is the limited number of points for TensorFlow. This is due to the fact that TensorFlow finishes a much smaller number of mega-batches than the other algorithms in the allocated time. In fact, it takes TensorFlow multiple days to process the number of batches -- 250 for Amazon-670k and 350 for Delicious-200k -- displayed in the figures. Overall, Adaptive SGD needs the fewest mega-batches to reach the highest accuracy -- which is higher than any other's algorithm -- on Amazon-670k. Moreover, the number of mega-batches decreases as more GPUs are added. This is somehow surprising because increasing the degree of parallelism is generally directly correlated with statistical efficiency. However, in Adaptive SGD -- unlike Elastic SGD and other previous methods -- the local replicas evolve independently based on dynamically assigned batches and model merging is a finely tuned operation that considers GPU heterogeneity. CROSSBOW requires fewer mega-batches to reach higher accuracy than Adaptive SGD when beginning processing on Delicious-200k. The reason is the more frequent model merging. However, the maximum accuracy it achieves is below that of Adaptive SGD because the local replicas start to diverge and they are never fully reconciled through model merging---unlike Adaptive SGD.

%%%%%%%%%%%%%%%%%%%%%%%%%%%%%%%%%%%%%
\begin{figure}[htbp]
\centering
\begin{subfigure}{.49\textwidth}
	\includegraphics[width=.49\linewidth]{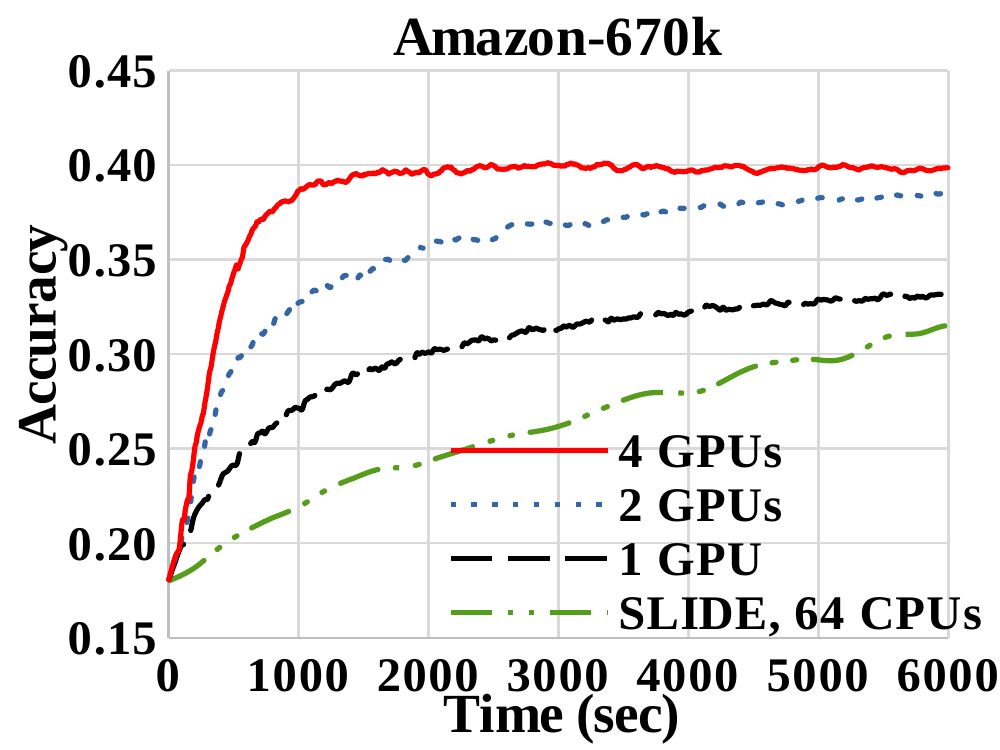}
  \includegraphics[width=.49\linewidth]{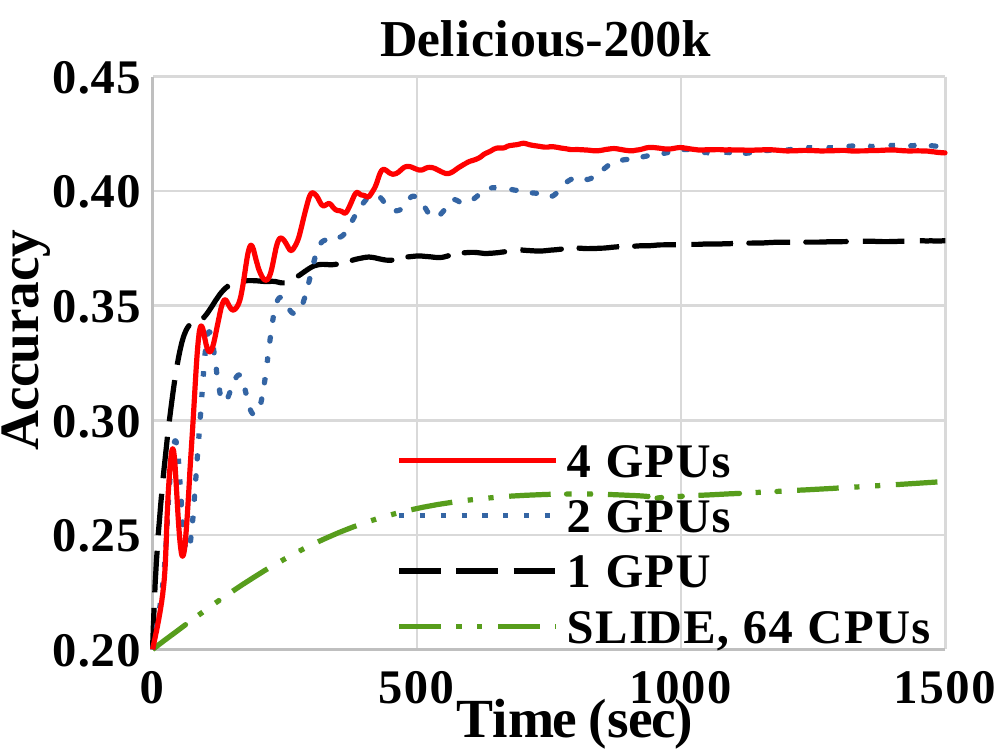}
\caption{Time-to-accuracy}
\label{fig:adaptive-gpus-time}
\end{subfigure}
\hfill
\begin{subfigure}{.49\textwidth}
	\includegraphics[width=.49\linewidth]{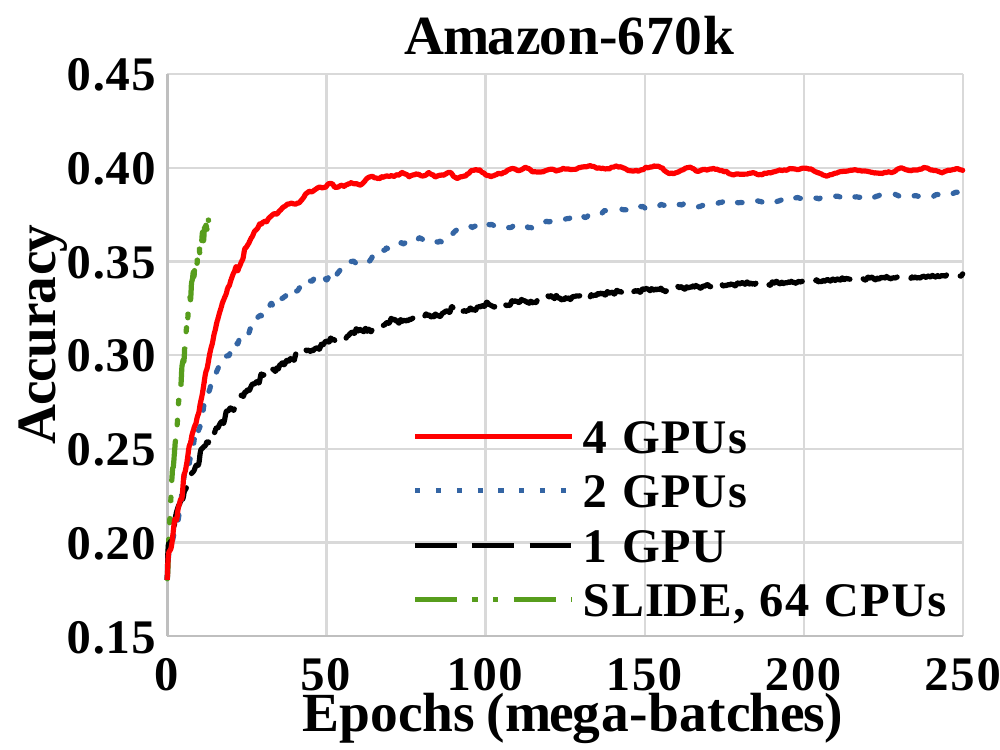}
  \includegraphics[width=.49\linewidth]{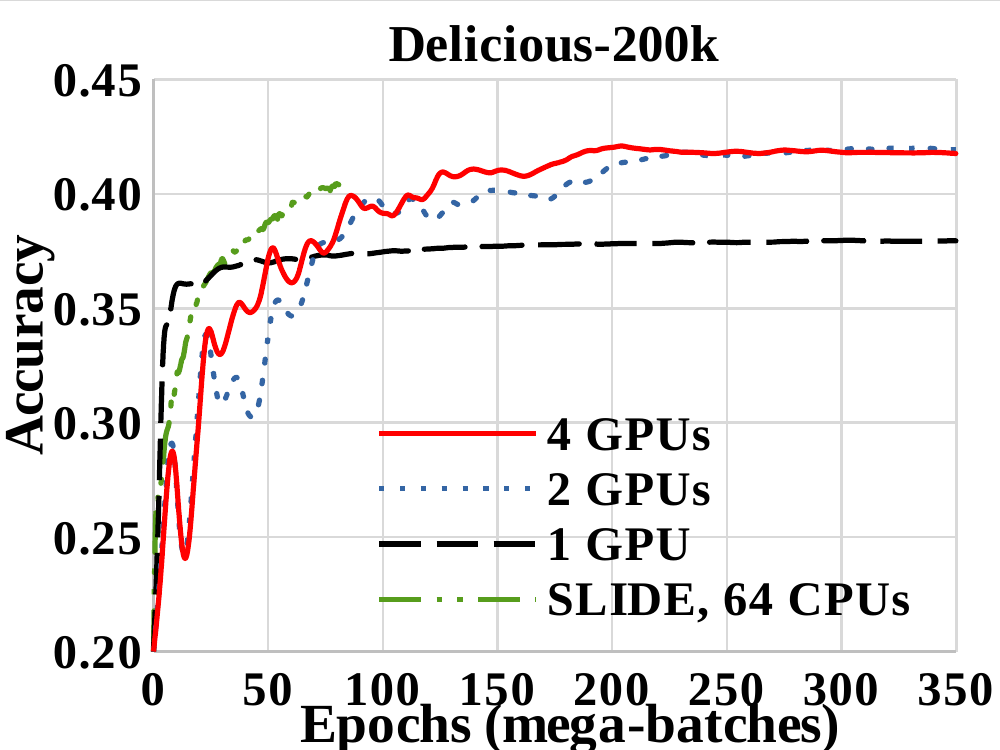}
\caption{Statistical efficiency}
\label{fig:adaptive-gpus-stateff}
\end{subfigure}
\caption{Time-to-accuracy (a) and statistical efficiency (b) comparison between Adaptive SGD and SLIDE.}
\end{figure}
%%%%%%%%%%%%%%%%%%%%%%%%%%%%%%%%%%%%%

%%%%%%%%%%%%%%%%%%%%%%%%%%%%%%%%%%%%%
\paragraph*{Scalability}
In order to analyze the training scalability of Adaptive SGD, we plot the time-to-accuracy (Figure~\ref{fig:adaptive-gpus-time}) and statistical efficiency (Figure~\ref{fig:adaptive-gpus-stateff}) as a function of the number of GPUs. We also include the curves for the optimized SLIDE algorithm as a CPU baseline in the figures. The benefit of using multiple GPUs is clear for the Amazon-670k dataset, where the accuracy on 4 GPUs is clearly better than on 2 and 1 GPU, respectively. This proves the importance of adaptive batch size scaling and normalized model merging. On the Delicious-200k dataset, the impact made by adding more GPUs is not so straightforward. The 4 GPUs configuration provides only a limited reduction in the time to reach the maximum accuracy over 2 GPUs. Moreover, these multi-GPU configurations require more epochs to ramp-up accuracy compared to the single GPU setting. Nonetheless, 4 GPUs always achieve the highest accuracy in the shortest time interval. Compared to SLIDE, all the Adaptive SGD GPU configurations -- including single GPU -- are superior to the optimized CPU algorithm. This is due to the much higher hardware efficiency of the GPU over CPU since the CPU statistical efficiency is net superior (Figure~\ref{fig:adaptive-gpus-stateff}). The reason is the higher number of model updates. Consequently, while specialized algorithms are valuable, they cannot easily outperform adequately tuned solutions on superior computing architectures.

%%%%%%%%%%%%%%%%%%%%%%%%%%%%%%%%%%%%%
\begin{figure}[htbp]
\centering
\begin{subfigure}{.49\textwidth}
	\includegraphics[width=.49\linewidth]{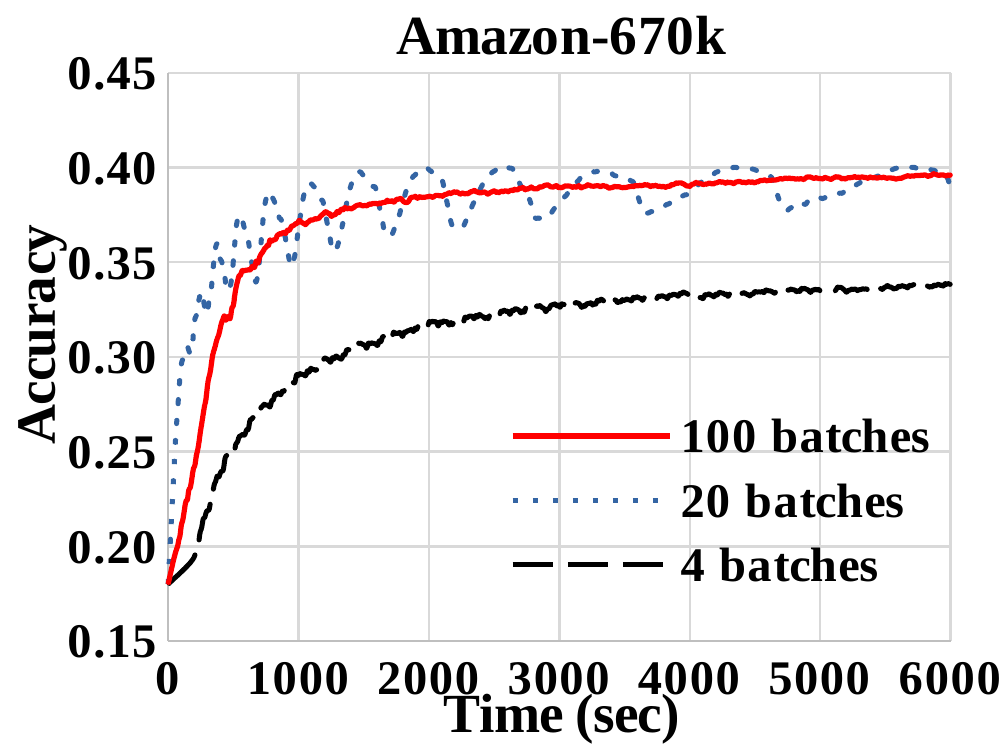}
  \includegraphics[width=.49\linewidth]{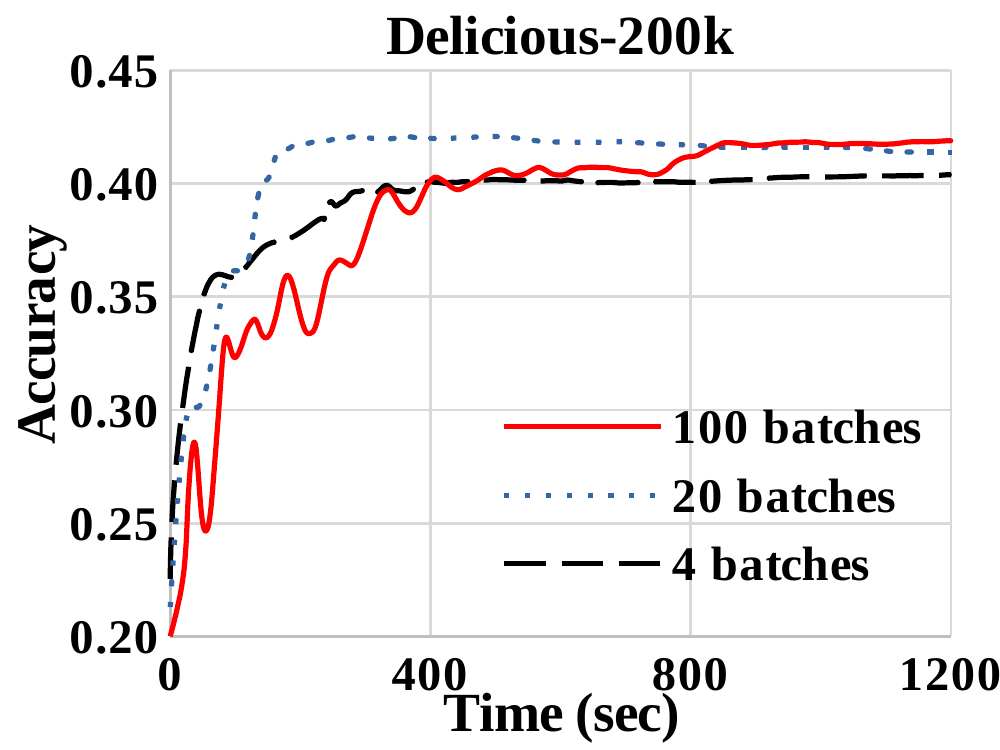}
\caption{Time-to-accuracy}
\label{fig:merge-freq-time}
\end{subfigure}
\hfill
\begin{subfigure}{.49\textwidth}
	\includegraphics[width=.49\linewidth]{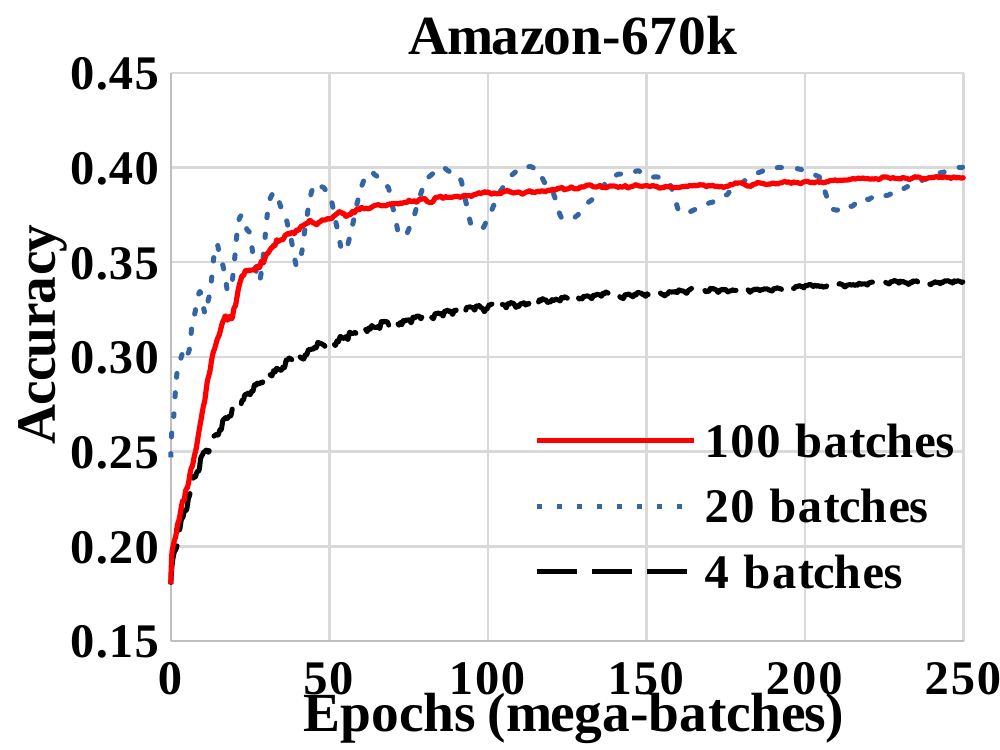}
  \includegraphics[width=.49\linewidth]{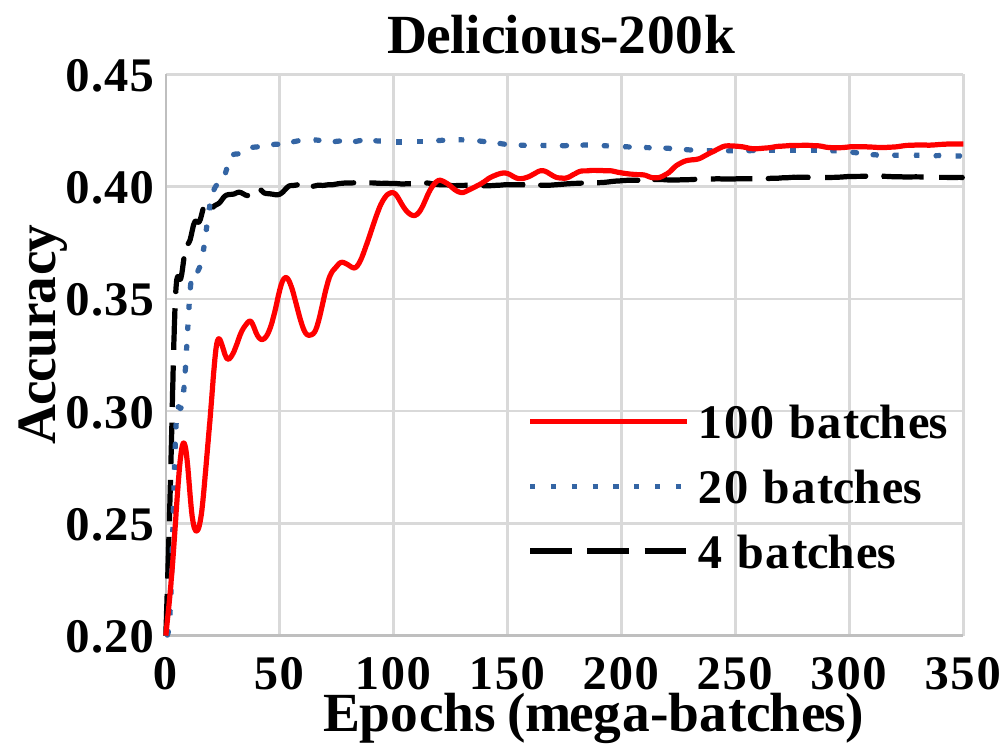}
\caption{Statistical efficiency}
\label{fig:merge-freq-iter}
\end{subfigure}
\caption{The effect of the mega-batch size (model merging frequency) on time-to-accuracy (a) and statistical efficiency (b) of Adaptive SGD for 4 GPUs.}
\end{figure}
%%%%%%%%%%%%%%%%%%%%%%%%%%%%%%%%%%%%%

%%%%%%%%%%%%%%%%%%%%%%%%%%%%%%%%%%%%%
\subsubsection{Micro-benchmarks for Parameter Sensitivity}

%%%%%%%%%%%%%%%%%%%%%%%%%%%%%%%%%%%%%
\paragraph*{Model merging frequency}
The size of a mega-batch determines the frequency of model merging. A large mega-batch allows the local replicas to evolve independently for a longer time period before they are merged. While this reduces the overhead incurred by merging, it also increases the relevance of merging since the local replicas may diverge. We depict the effect of the mega-batch size -- as a number of batches -- on time-to-accuracy (Figure~\ref{fig:merge-freq-time}) and statistical efficiency (Figure~\ref{fig:merge-freq-iter}). Merging after a mega-batch of 4 batches on 4 GPUs corresponds to gradient aggregation, while the other alternatives illustrate the benefits of the proposed Adaptive SGD. As expected, with a larger mega-batch, both the time-to-accuracy and the number of epochs to achieve a certain level of accuracy increase. However, the improvement over gradient aggregation is evident. Due to batch size adaptation and perturbation, Adaptive SGD explores the optimization space more thoroughly and achieves higher accuracy---impossible to reach by gradient aggregation. While the optimal choice of model merging frequency depends on the dataset, a mega-batch of 20 or more batches performs well in practice. We use a value of 100 throughout the experiments in order to allow for increased adaptivity.

%%%%%%%%%%%%%%%%%%%%%%%%%%%%%%%%%%%%%
\begin{figure}[htbp]
\centering
\begin{subfigure}{.49\textwidth}
	\includegraphics[width=.48\textwidth]{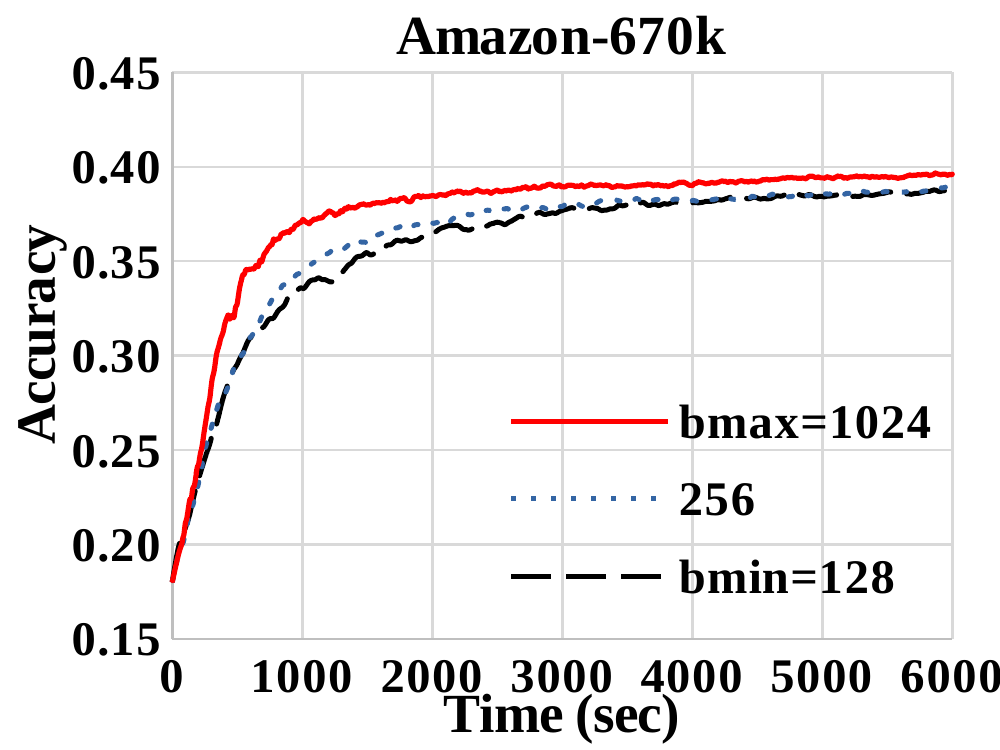}
	\includegraphics[width=.48\textwidth]{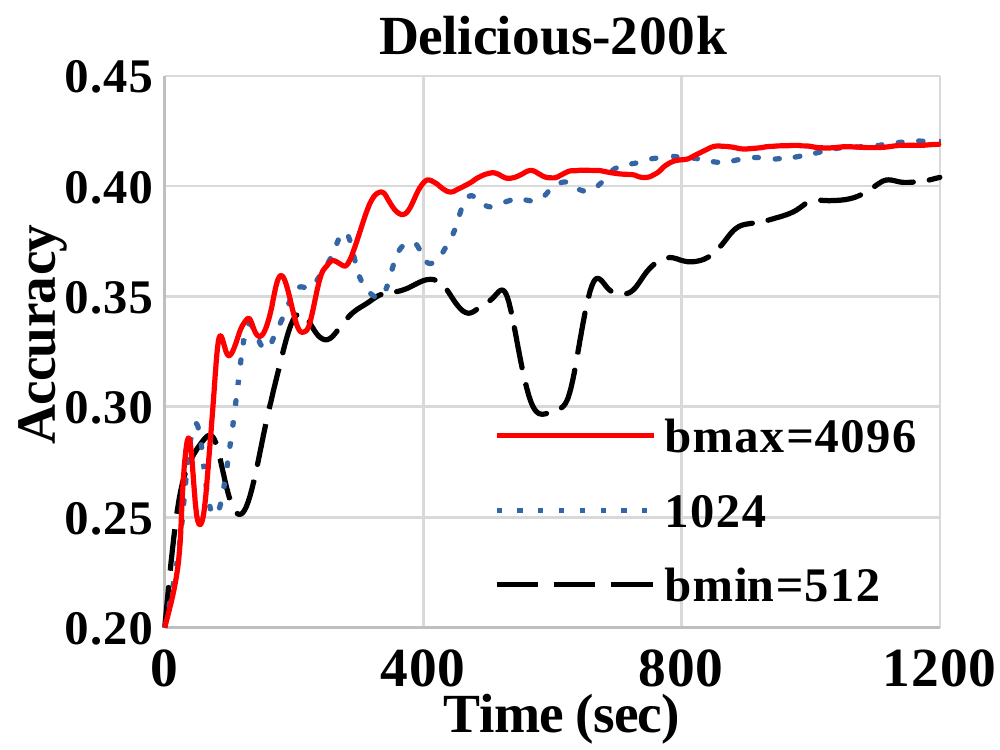}
\caption{Initial batch size}
\label{fig:initial-batch}
\end{subfigure}
\hfill
\begin{subfigure}{.49\textwidth}
	\includegraphics[width=.48\textwidth]{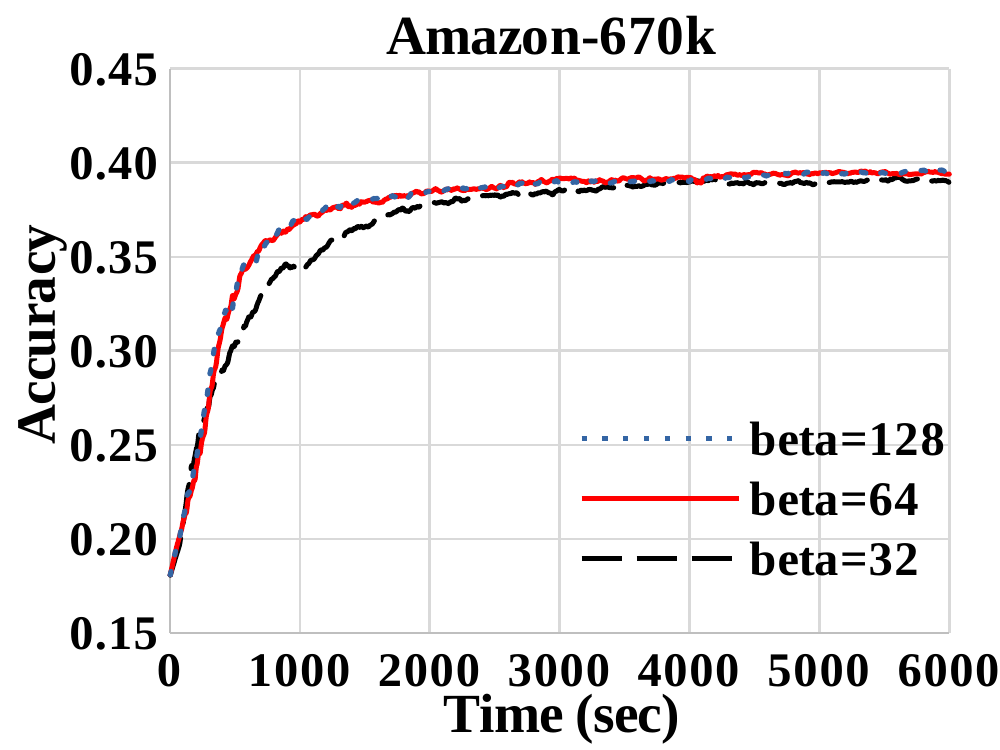}
	\includegraphics[width=.48\textwidth]{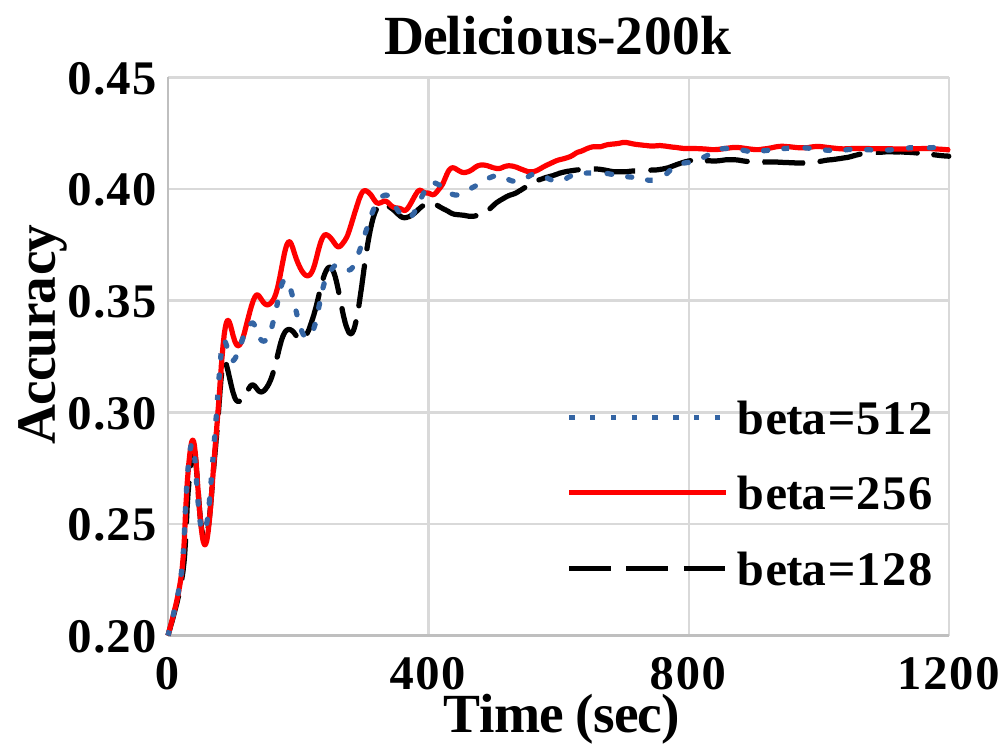}
\caption{Batch size scaling factor $\beta$}
\label{fig:batch-size-beta}
\end{subfigure}
\caption{The effect of the initial batch size (a) and the batch size scaling parameter $\beta$ (b) on the time-to-accuracy of Adaptive SGD for 4 GPUs.}
\end{figure}
%%%%%%%%%%%%%%%%%%%%%%%%%%%%%%%%%%%%%

%%%%%%%%%%%%%%%%%%%%%%%%%%%%%%%%%%%%%
\paragraph*{Initial batch size}
While the batch size stays unchanged for the non-adaptive algorithms, in Adaptive SGD the batch size corresponding to a GPU evolves based on its speed relative to the other GPUs. Since the change for every mega-batch is limited, the initial batch size may impact the early performance of Adaptive SGD. This is exactly what Figure~\ref{fig:initial-batch} shows. In the initial stages, the accuracy is higher when the batch size is starting from the maximum allowable value $b_{\textit{max}}$ in Algorithm~\ref{alg:batch-size-scale}. Moreover, the smaller the initial value is, the longer it takes to reach the same level of accuracy. Since the linear scaling rule is applied to determine the learning rate, the reason for this delay is exclusively the overhead incurred by smaller batches. Also, larger batches allow for more accurate estimation of a GPU's speed. This is evident from the figure corresponding to the Delicious-200k dataset where the accuracy for $b_{\textit{min}}$ experiences heavy oscillations. Given these, we set the initial batch size to $b_{\textit{max}}$.

%%%%%%%%%%%%%%%%%%%%%%%%%%%%%%%%%%%%%
\paragraph*{Batch size scaling factor $\beta$}
The scaling factor $\beta$ controls how large is the increase or decrease of the batch size in the linear scaling algorithm (Algorithm~\ref{alg:batch-size-scale}). The larger $\beta$ is, the higher the change in the batch size. We take half of the minimum allowable batch size $b_{\textit{min}}$ as the baseline value for $\beta$. We consider two additional values -- one smaller and the other larger -- for every dataset. Their impact on time-to-accuracy is depicted in Figure~\ref{fig:batch-size-beta}. We observe a small difference among these alternatives, with a slight advantage for the larger values. When $\beta$ is too small, the change in the batch size is insufficient to quickly compensate for the speed difference among GPUs. Moreover, a too large $\beta$ is prone to high oscillations in accuracy---the case for Delicious-200k. Thus, we settle for a $\beta$ value equal to half of the minimum possible batch size since it can be readily determined.

%%%%%%%%%%%%%%%%%%%%%%%%%%%%%%%%%%%%%
\begin{figure}[htbp]
\centering
\begin{subfigure}{.49\textwidth}
	\includegraphics[width=.48\textwidth]{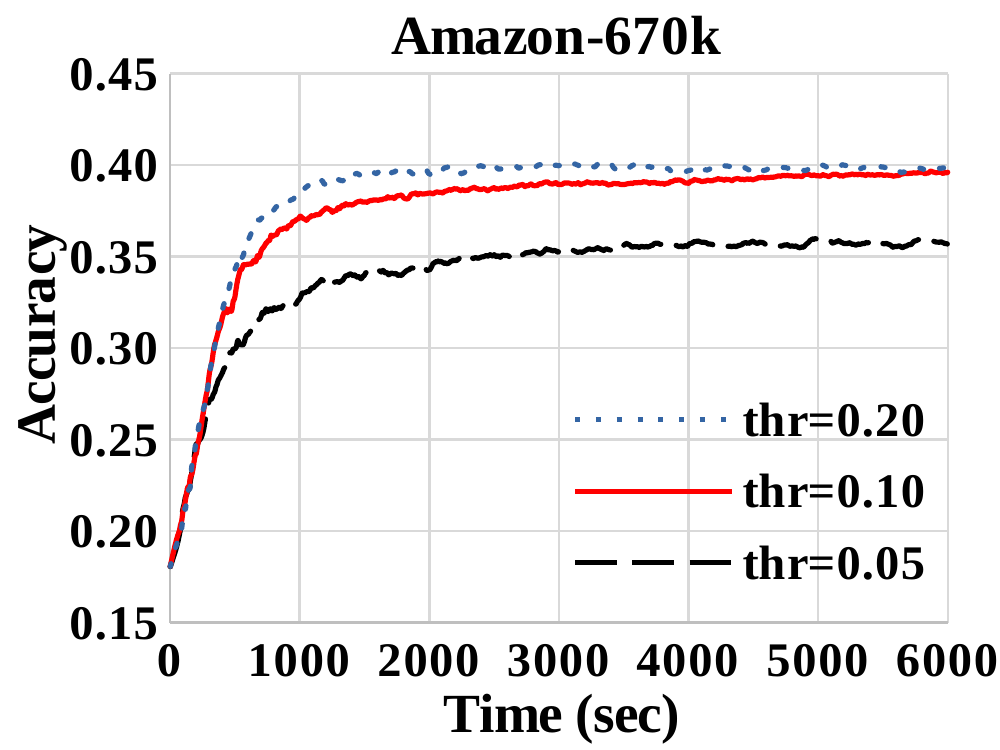}
	\includegraphics[width=.48\textwidth]{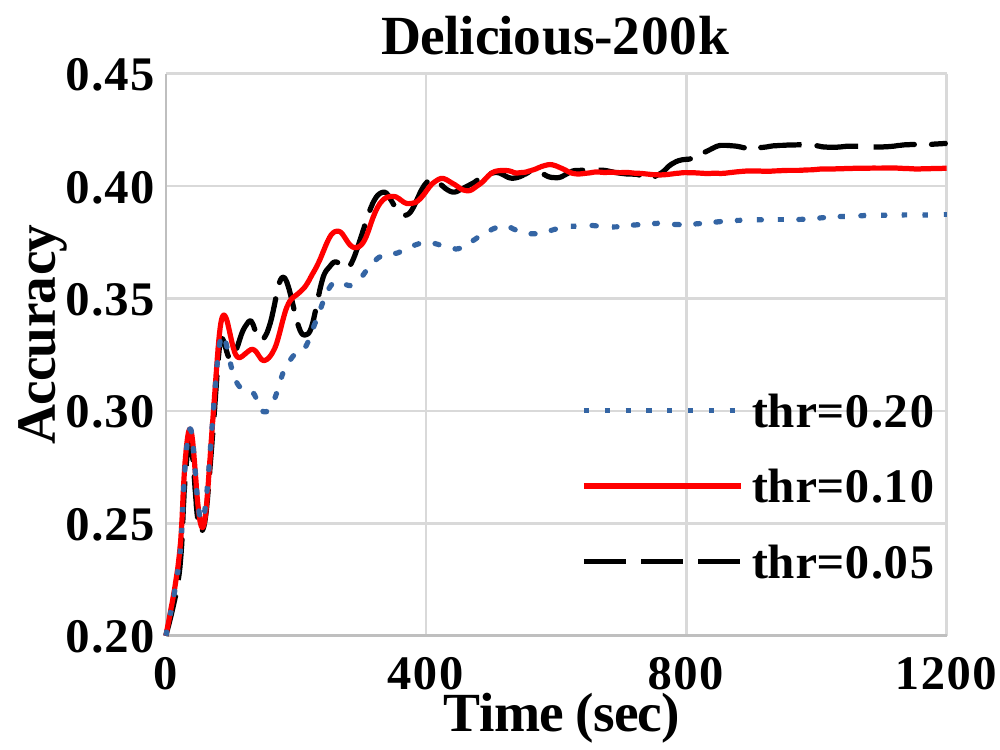}
\caption{Perturbation threshold $\textit{pert}_{\textit{thr}}$}
\label{fig:l2-norm}
\end{subfigure}
\hfill
\begin{subfigure}{.49\textwidth}
	\includegraphics[width=.48\textwidth]{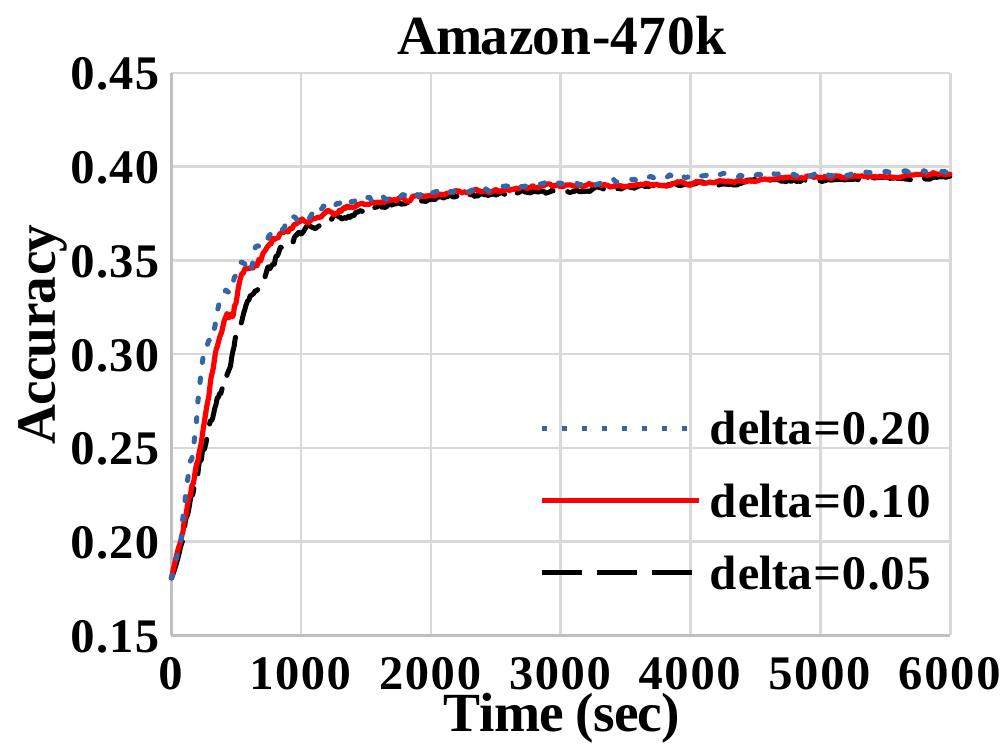}
	\includegraphics[width=.48\textwidth]{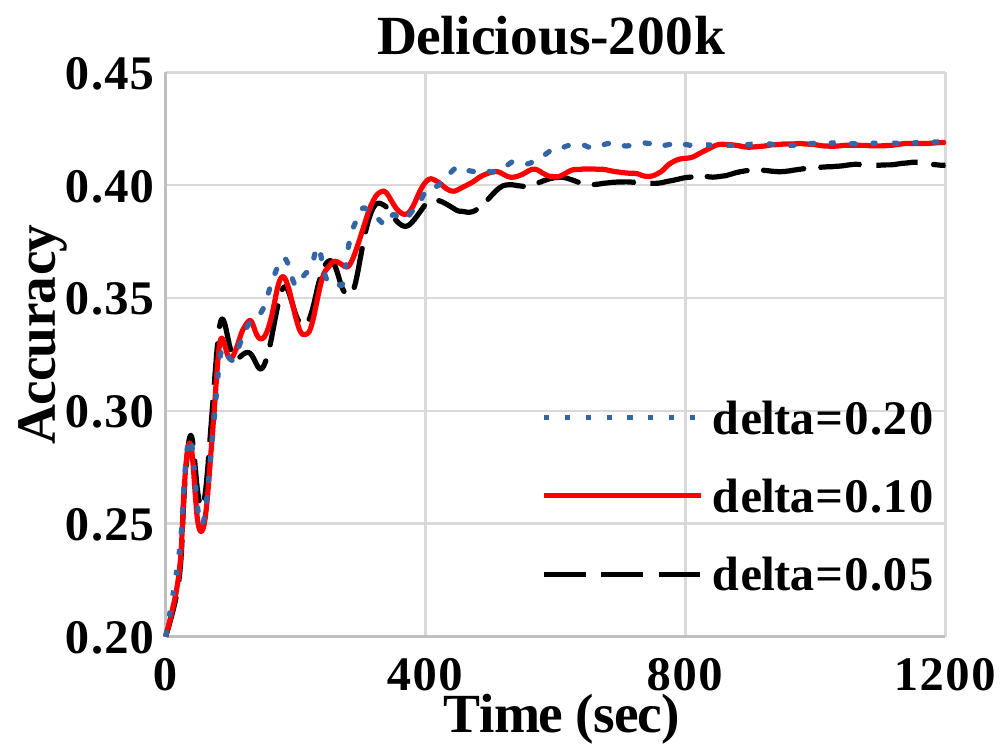}
\caption{Perturbation factor $\delta$}
\label{fig:delta}
\end{subfigure}
\caption{The effect of the perturbation threshold $\textit{pert}_{\textit{thr}}$ (a) and perturbation factor $\delta$ (b) on the time-to-accuracy of Adaptive SGD for 4 GPUs.}
\end{figure}
%%%%%%%%%%%%%%%%%%%%%%%%%%%%%%%%%%%%%

%%%%%%%%%%%%%%%%%%%%%%%%%%%%%%%%%%%%%
\paragraph*{Perturbation threshold $\textit{pert}_{\textit{thr}}$}
The parameter $\textit{pert}_{\textit{thr}}$ is an indicator of how regularized a model replica is. It controls the frequency at which perturbation is added to the weighted model averaging in Algorithm~\ref{alg:norm-model-merge}. The higher the threshold is, the more perturbation is added to model merging. Perturbation increases the weight of the most updated replica, pulling the global model into that direction. Figure~\ref{fig:l2-norm} depicts the accuracy of Adaptive SGD for three different thresholds. On Amazon-670k, a small threshold of 0.05 blocks perturbation, resulting in less accuracy. On Delicious-200k, the opposite effect is observed. The highest accuracy is obtained with a threshold of 0.05. The trend of the global model's L2-norm is the main cause for this phenomenon. On Amazon-670k, the L2-norm keeps increasing with the increase in accuracy, while on Delicious-200k the opposite holds. Since a median threshold of 0.10 is less sensitive to these variations, we adopt this value throughout the experiments.

%%%%%%%%%%%%%%%%%%%%%%%%%%%%%%%%%%%%%
\paragraph*{Perturbation factor $\delta$}
The degree of perturbation added to model merging is determined by the value of parameter $\delta$ in Algorithm~\ref{alg:norm-model-merge}. The larger this value is, the more dominant the most updated model replica is. However, in order to limit the denormalization induced by perturbation, the maximum value of $\delta$ has to be restricted. Figure~\ref{fig:delta} shows the effect of three different $\delta$ values on accuracy. There is a small difference among them since only the weights of two local replicas are modified. Thus, in order to limit the number of parameter values used in Adaptive SGD, we opt for a $\delta$ of 0.10---as is the case for the threshold $\textit{pert}_{\textit{thr}}$.

%%%%%%%%%%%%%%%%%%%%%%%%%%%%%%%%%%%%%
\begin{figure}[htbp]
\centering
\begin{subfigure}{.49\textwidth}
	\includegraphics[width=.48\linewidth]{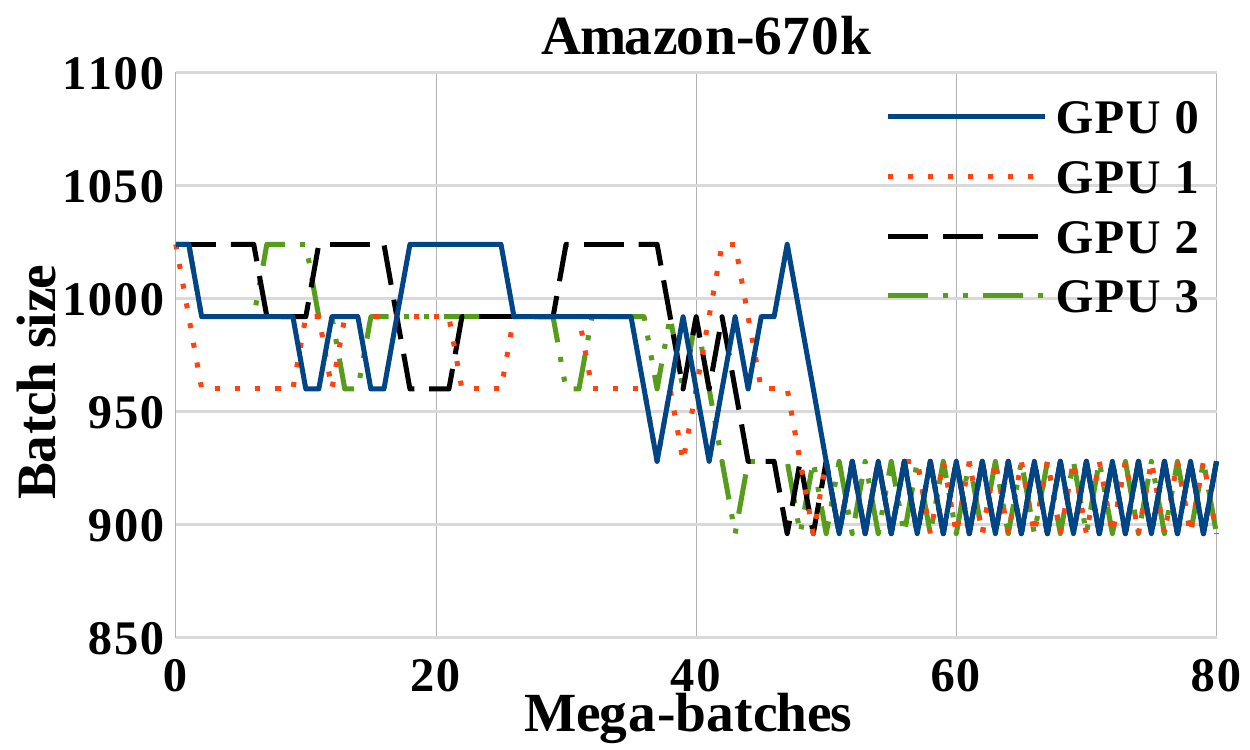}
	\includegraphics[width=.48\linewidth]{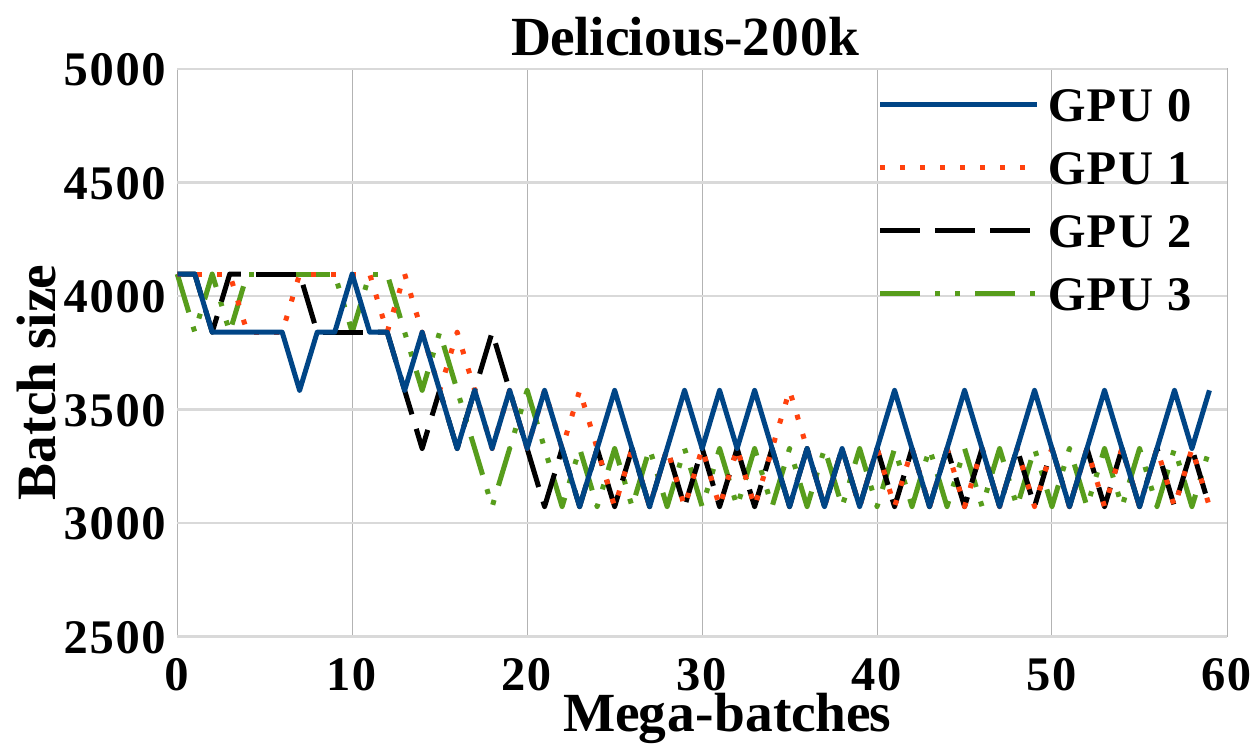}
\caption{Batch size scaling}
\label{fig:batch_size}
\end{subfigure}
\hfill
\begin{subfigure}{.49\textwidth}
	\includegraphics[width=.48\linewidth]{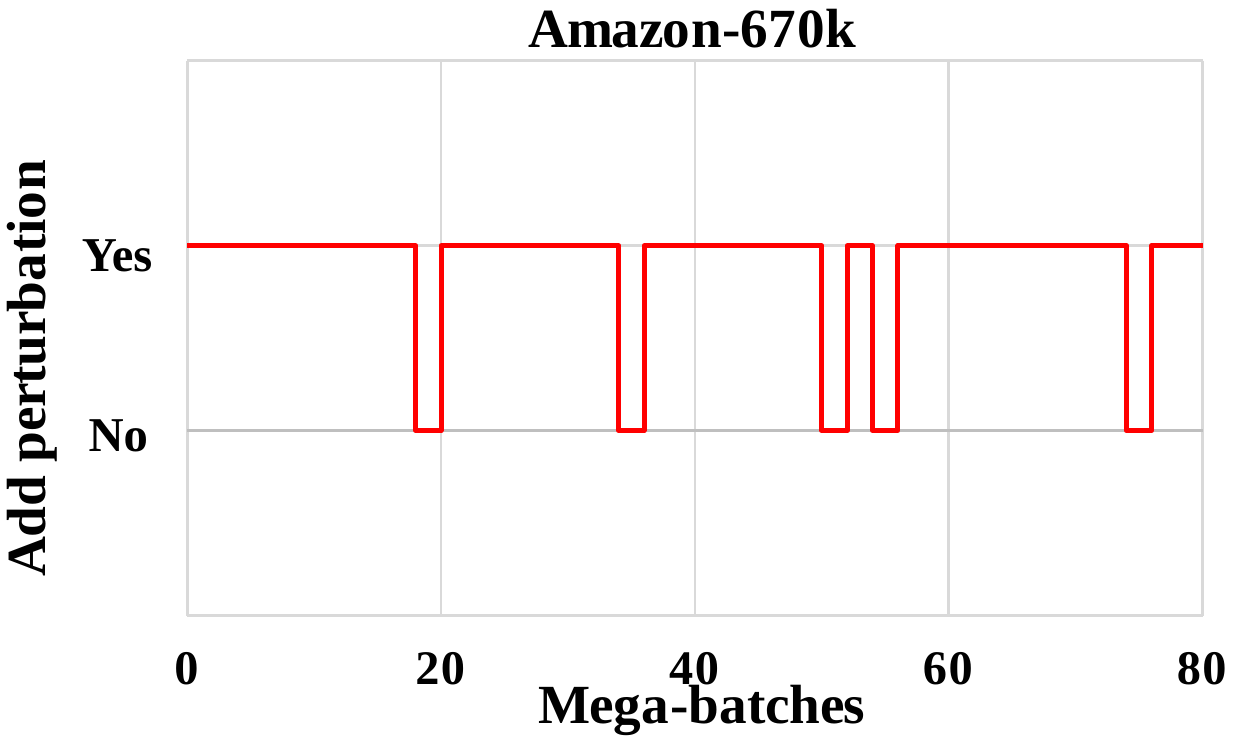}
	\includegraphics[width=.48\linewidth]{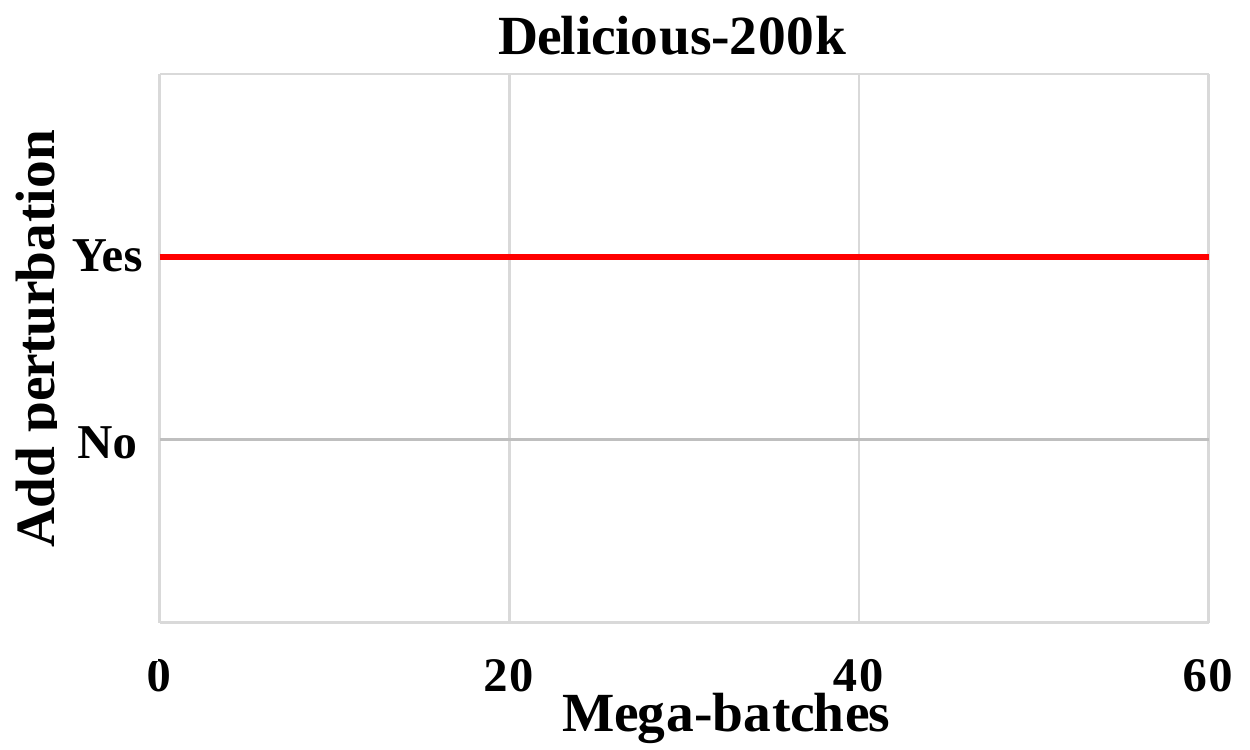}
\caption{Perturbation frequency}
\label{fig:weight_norm}
\end{subfigure}
\caption{The batch size of a GPU is adaptively scaled (a) and perturbation is frequently added to normalized model merging (b) after processing every mega-batch.}
\end{figure}
%%%%%%%%%%%%%%%%%%%%%%%%%%%%%%%%%%%%%

%%%%%%%%%%%%%%%%%%%%%%%%%%%%%%%%%%%%%
\paragraph*{Do batch size scaling and perturbation activate in Adaptive SGD?}
Although batch size scaling and perturbed model merging are part of the Adaptive SGD algorithm, their activation is conditioned by certain triggers. Batch scaling happens only when the GPUs process a different number of replica updates, while perturbation is added to model merging only when all the replicas are well-regularized. In order to quantify how often these conditions are met during training, we plot the evolution of the batch size for every GPU in Figure~\ref{fig:batch_size} and the activation frequency of perturbation in Figure~\ref{fig:weight_norm}. We observe how the batch size changes across GPUs after every mega-batch. The batch sizes are initialized with the maximum allowable value and fluctuate until they converge to a limited range in which the number of updates to the local replicas stabilize. At this point, all the GPUs perform a synchronized operation that is optimal for training. The perturbation frequency is controlled by the threshold $\textit{pert}_{\textit{thr}}$---set to 0.10 following an empirical study. From Figure~\ref{fig:weight_norm}, we see that perturbation is added to merging with a very high frequency, meaning that the local replicas are well-regularized. These results confirm that the novel characteristics of Adaptive SGD are the source for its improved performance.

%%%%%%%%%%%%%%%%%%%%%%%%%%%%%%%%%%%%%
\subsection{Summary}\label{ssec:experiments:discussion}

Based on the presented results, we can answer the questions raised at the beginning of this section:
\begin{itemize}[leftmargin=*,noitemsep,nolistsep]
\item Due to the careful handling of heterogeneity -- which allows a more thorough exploration of the optimization space -- Adaptive SGD outperforms all the competitors in time-to-accuracy. In fact, Adaptive SGD achieves the highest accuracy among all the algorithms. In terms of statistical efficiency, SLIDE requires fewer epochs to achieve a certain level of accuracy due to the larger number of model updates performed across multiple CPU threads. However, this requires substantially longer time than Adaptive SGD.
\item As we increase the number of GPUs, Adaptive SGD exhibits both faster time-to-accuracy -- which is expected -- as well as less epochs and higher overall accuracy---which cannot be reached by any of the other methods.
\item A large mega-batch size increases the time-to-accuracy only in the initial epochs of Adaptive SGD. As the algorithm progresses, even a very large mega-batch consisting of 100 batches achieves optimal time-to-accuracy. Thus, this is the value used in the experiments.
\item Although Adaptive SGD has multiple parameters, their values can be set mostly independently of the training dataset. Specifically, the initial batch size is set to the maximum possible value that fits in the GPU memory, while the batch size scaling factor to ${1/16}^{\text{th}}$ of the initial batch size. Additionally, 0.10 is a good value both for the perturbation threshold and the perturbation factor, reducing the overall number of distinct parameter values.
\item Adaptive batch size scaling and normalized model merging with perturbation -- the two identifying characteristics of Adaptive SGD -- are frequently invoked during execution, which confirms both the need to carefully consider heterogeneity and that they are the reason for the superior Adaptive SGD performance.
\end{itemize}

%%%%%%%%%%%%%%%%%%%%%%%%%%%%%%%%%%%%%%%%%%%%%%%%%%%%%%%%%%%%%%%%%
%\input{conclusions}
\section{CONCLUSIONS}\label{sec:conclusions}

In this paper, we introduce Adaptive SGD, an adaptive elastic model averaging stochastic gradient descent algorithm for heterogeneous multi-GPUs. Adaptive SGD addresses the challenges raised by parallel training large machine learning models on sparse data with dynamic scheduling, adaptive batch size scaling, and normalized model merging. Instead of statically assigning batches to GPUs, dynamically scheduled batches are allocated based on the relative GPU processing speed. Since this can lead to a different number of model updates across GPUs, batch size scaling assigns larger batches to the faster GPUs and smaller batches to the slower ones, with the goal to arrive at a steady state in which all the GPUs perform the same number of model updates. The batch sizes are continuously updated following a linear function that quantifies the deviation from the expected number of updates, while guaranteeing a minimum degree of GPU utilization and imposing strict bounds on model replica staleness. Normalized model merging computes optimal weights for every GPU based on the assigned batches. The underlying principle is to prioritize the replicas updated more frequently and with gradients derived from larger batch sizes. We provide an open-source implementation of Adaptive SGD and compare its performance against four existing methods. The results show that Adaptive SGD outperforms all the other solutions in time-to-accuracy and confirm its scalability with the number of GPUs. Moreover, we empirically determine optimal values for all of Adaptive SGD's parameters.

%%%%%%%%%%%%%%%%%%%%%%%%%%%%%%%%%%%%%%%%%%%%%%%%%%%%%%%
\paragraph*{Acknowledgments}
This work is supported by a U.S. Department of Energy Early Career Award (DOE Career), the Office of Advanced Scientific Computing Research (ASCR), Office of Science, of the U.S. Department of Energy under Contract No. DE-AC02-05CH11231, and uses resources of the National Energy Research Scientific Computing Center (NERSC).

%%%%%%%%%%%%%%%%%%%%%%%%%%%%%%%%%%%%%%%%%%%%%%%%%%%%%%
\bibliographystyle{abbrv}
%\bibliography{biblio}

\begin{thebibliography}{10}

\bibitem{tensorflow}
{M. Abadi et al.}
\newblock {TensorFlow: {A} System for Large-Scale Machine Learning}.
\newblock In {\em OSDI 2016}, pages 265--283.

\bibitem{lock-free-sgd:ipdps-2021}
K.~Backstrom, I.~Walulya, M.~Papatriantafilou, and P.~Tsigas.
\newblock {Consistent Lock-free Parallel Stochastic Gradient Descent for Fast
  and Stable Convergence}.
\newblock In {\em {IPDPS 2021}}.

\bibitem{couple-adaptive-batch-size-learn-rate}
L.~Balles, J.~Romero, and P.~Hennig.
\newblock {Coupling Adaptive Batch Sizes with Learning Rates}.
\newblock {\em CoRR}, arXiv/1612.05086v2, 2017.

\bibitem{bertsekas:igd}
D.~P. Bertsekas.
\newblock {Incremental Gradient, Subgradient, and Proximal Methods for Convex
  Optimization: A Survey}.
\newblock MIT 2010.

\bibitem{Xclassification-dataset-repo}
K.~Bhatia, K.~Dahiya, H.~Jain, A.~Mittal, Y.~Prabhu, and M.~Varma.
\newblock {The Extreme Classification Repository: Multi-label Datasets and
  Code}.
\newblock \url{http://manikvarma.org/downloads/XC/XMLRepository.html}, 2016.

\bibitem{deep-nets-sgd}
L.~Bottou.
\newblock {\em {Neural Networks: Tricks of the Trade}}.
\newblock Springer, 2012.

\bibitem{bottou:optim-methods-scale-ml}
L.~Bottou, F.~Curtis, and J.~Nocedal.
\newblock {Optimization Methods for Large-Scale Machine Learning}.
\newblock {\em SIAM Review}, 60(2):223--311, 2018.

\bibitem{adaptive-batch-size-variance:mp-2012}
R.~H. Byrd, G.~M. Chin, J.~Nocedal, and Y.~Wu.
\newblock {Sample Size Selection in Optimization Methods for Machine Learning}.
\newblock {\em Mathematical Programming}, 134(1):127--155, 2012.

\bibitem{slide:arxiv}
B.~Chen, T.~Medini, and A.~Shrivastava.
\newblock {{SLIDE} : In Defense of Smart Algorithms over Hardware Acceleration
  for Large-Scale Deep Learning Systems}.
\newblock {\em CoRR}, arXiv/1903.03129, 2019.

\bibitem{sync-vs-asynch-sgd}
J.~Chen, R.~Monga, S.~Bengio, and R.~J{\'{o}}zefowicz.
\newblock {Revisiting Distributed Synchronous {SGD}}.
\newblock {\em CoRR}, arXiv/1604.00981, 2016.

\bibitem{mxnet}
T.~Chen, M.~Li, Y.~Li, M.~Lin, N.~Wang, M.~Wang, T.~Xiao, B.~Xu, C.~Zhang, and
  Z.~Zhang.
\newblock {MXNet: {A} Flexible and Efficient Machine Learning Library for
  Heterogeneous Distributed Systems}.
\newblock {\em CoRR}, arXiv/1512.01274, 2015.

\bibitem{geeps}
H.~Cui, H.~Zhang, G.~R. Ganger, P.~B. Gibbons, and E.~P. Xing.
\newblock {GeePS: Scalable Deep Learning on Distributed GPUs with a
  GPU-specialized Parameter Server}.
\newblock In {\em {EuroSys 2016}}.

\bibitem{deep-xml}
K.~Dahiya, D.~Saini, A.~Mittal, A.~Shaw, K.~Dave, A.~Soni, H.~Jain, S.~Agarwal,
  and M.~Varma.
\newblock {DeepXML: A Deep Extreme Multi-label Learning Framework Applied to
  Short Text Documents}.
\newblock In {\em WSDM 2021}.

\bibitem{adaptive-batch-size-variance:aistats-2017}
S.~De, A.~Yadav, D.~Jacobs, and T.~Goldstein.
\newblock {Automated Inference with Adaptive Batches}.
\newblock In {\em AISTATS 2017}.

\bibitem{saga}
A.~Defazio, F.~Bach, and S.~Lacoste-Julien.
\newblock {SAGA: A Fast Incremental Gradient Method with Support for
  Non-strongly Convex Composite Objectives}.
\newblock In {\em NIPS 2014}, pages 1646--1654.

\bibitem{finance-sgd}
M.~Dixon, D.~Klabjan, and J.~Bang.
\newblock {Classification-based Financial Markets Prediction using Deep Neural
  Networks}.
\newblock {\em CoRR}, arXiv/1603.08604, 2016.

\bibitem{dnn-par+dis-survey}
A.~Farkas, G.~Kertesz, and R.~Lovas.
\newblock {Parallel and Distributed Training of Deep Neural Networks: A Brief
  Overview}.
\newblock In {\em INES 2020}, pages 165--170.

\bibitem{adaptive-batch-size-constant}
M.~P. Friedlander and M.~Schmidt.
\newblock {Hybrid Deterministic-Stochastic Methods for Data Fitting}.
\newblock {\em SIAM Journal on Scientific Computing}, 34(3):1380--1405, 2012.

\bibitem{facebook:large-batches}
P.~Goyal, L.~Wesolowski, P.~Dollar, A.~Kyrola, R.~Girshick, A.~Tulloch,
  P.~Noordhuis, Y.~Jia, and K.~He.
\newblock {Accurate, Large Minibatch SGD: Training ImageNet in 1 Hour}.
\newblock {\em CoRR}, arXiv/1706.02677v2, 2018.

\bibitem{omnivore}
S.~Hadjis, C.~Zhang, I.~Mitliagkas, and C.~R{\'{e}}.
\newblock {Omnivore: An Optimizer for Multi-device Deep Learning on CPUs and
  GPUs}.
\newblock {\em CoRR}, arXiv/1606.04487, 2016.

\bibitem{speech-sgd}
{G. Hinton et al.}
\newblock {Deep Neural Networks for Acoustic Modeling in Speech Recognition}.
\newblock {\em IEEE Signal Processing Magazine}, 29:82--97, 2012.

\bibitem{stale-synch-ps}
Q.~Ho, J.~Cipar, H.~Cui, S.~Lee, J.~K. Kim, P.~B. Gibbons, G.~A. Gibson,
  G.~Ganger, and E.~P. Xing.
\newblock {More Effective Distributed ML via a Stale Synchronous Parallel
  Parameter Server}.
\newblock In {\em NIPS 2013}, pages 1223--1231.

\bibitem{flex-ps}
Y.~Huang, T.~Jin, Y.~Wu, Z.~Cai, X.~Yan, F.~Yang, J.~Li, Y.~Guo, and J.~Cheng.
\newblock {FlexPS: Flexible Parallelism Control in Parameter Server
  Architecture}.
\newblock {\em PVLDB}, 11(5):566--579, 2018.

\bibitem{hetero-ps}
J.~Jiang, B.~Cui, C.~Zhang, and L.~Yu.
\newblock {Heterogeneity-aware Distributed Parameter Servers}.
\newblock In {\em {SIGMOD 2017}}, pages 463--478.

\bibitem{light-xml}
T.~Jiang, D.~Wang, L.~Sun, H.~Yang, Z.~Zhao, and F.~Zhuang.
\newblock {LightXML: Transformer with Dynamic Negative Sampling for
  High-Performance Extreme Multi-label Text Classification}.
\newblock In {\em AAAI 2021}.

\bibitem{svrg}
R.~Johnson and T.~Zhang.
\newblock {Accelerating Stochastic Gradient Descent Using Predictive Variance
  Reduction}.
\newblock In {\em NIPS 2013}, pages 315--323.

\bibitem{crossbow}
A.~Koliousis, P.~Watcharapichat, M.~Weidlich, L.~Mai, P.~Costa, and
  P.~Pietzuch.
\newblock {CROSSBOW: Scaling Deep Learning with Small Batch Sizes on Multi-GPU
  Servers}.
\newblock {\em PVLDB}, 12(11):1399--1413, 2019.

\bibitem{one-weird-trick}
A.~Krizhevsky.
\newblock {One Weird Trick for Parallelizing Convolutional Neural Networks}.
\newblock {\em CoRR}, arXiv/1404.5997, 2014.

\bibitem{parameter-server}
M.~Li, L.~Zhou, Z.~Yang, A.~Li, F.~Xia, D.~G. Andersen, and A.~Smola.
\newblock {Scaling Distributed Machine Learning with the Parameter Server}.
\newblock In {\em OSDI 2014}.

\bibitem{asynch-elastic-sgd}
X.~Lian, W.~Zhang, C.~Zhang, and J.~Liu.
\newblock {Asynchronous Decentralized Parallel Stochastic Gradient Descent}.
\newblock In {\em ICML 2018}, pages 3043--3052.

\bibitem{hetero-sgd:arxiv}
Y.~Ma and F.~Rusu.
\newblock {Heterogeneous CPU+GPU Stochastic Gradient Descent Algorithms}.
\newblock {\em CoRR}, arXiv/2004.08771, 2020.

\bibitem{bgd-vs-sgd:ipdps-2019}
Y.~Ma, F.~Rusu, and M.~Torres.
\newblock {Stochastic Gradient Descent on Modern Hardware: Multi-core CPU or
  GPU? Synchronous or Asynchronous?}
\newblock In {\em {IPDPS 2019}}.

\bibitem{hetero-sgd:ipdpsw-2021}
Y.~Ma, F.~Rusu, K.~Wu, and A.~Sim.
\newblock {Adaptive Stochastic Gradient Descent for Deep Learning on
  Heterogeneous CPU+GPU Architectures}.
\newblock In {\em {IPDPSW 2021}}.

\bibitem{revisit-small-batch}
D.~Masters and C.~Luschi.
\newblock {Revisiting Small Batch Training for Deep Neural Networks}.
\newblock {\em CoRR}, arXiv/1804.07612, 2018.

\bibitem{large-batch-empirical}
S.~McCandlish, J.~Kaplan, and D.~Amodei.
\newblock {An Empirical Model of Large-Batch Training}.
\newblock {\em CoRR}, arXiv/1812.06162, 2018.

\bibitem{cerebro}
S.~Nakandala, Y.~Zhang, and A.~Kumar.
\newblock {Cerebro: A Data System for Optimized Deep Learning Model Selection}.
\newblock {\em PVLDB}, 13(11):2159--2173, 2020.

\bibitem{hogwild}
F.~Niu, B.~Recht, C.~R{\'e}, and S.~J. Wright.
\newblock {Hogwild: A Lock-Free Approach to Parallelizing Stochastic Gradient
  Descent}.
\newblock In {\em NIPS 2011}.

\bibitem{pytorch}
{A. Paszke et al.}
\newblock {PyTorch: An Imperative Style, High-Performance Deep Learning
  Library}.
\newblock In {\em NeurIPS 2019}, pages 8024--8035.

\bibitem{hogwild-disk}
C.~Qin, M.~Torres, and F.~Rusu.
\newblock {Scalable Asynchronous Gradient Descent Optimization for Out-of-Core
  Models}.
\newblock {\em PVLDB}, 10(10):986--997, 2017.

\bibitem{combustion-dnn}
T.~Ren, M.~F. Modest, A.~Fateev, G.~Sutton, W.~Zhao, and F.~Rusu.
\newblock {Machine Learning Applied to Retrieval of Temperature and
  Concentration Distributions from Infrared Emission Measurements}.
\newblock {\em Applied Energy}, 252(113448), 2019.

\bibitem{sgd:survey}
S.~Ruder.
\newblock {An Overview of Gradient Descent Optimization Algorithms}.
\newblock {\em CoRR}, arXiv/1609.04747v2, 2017.

\bibitem{hogbatch}
S.~Sallinen, N.~Satish, M.~Smelyanskiy, S.~Sury, and C.~R{\'e}.
\newblock {High Performance Parallel Stochastic Gradient Descent in Shared
  Memory}.
\newblock In {\em IPDPS 2016}.

\bibitem{cntk}
F.~Seide and A.~Agarwal.
\newblock {CNTK: Microsoft's Open-Source Deep-Learning Toolkit}.
\newblock In {\em {KDD 2016}}, pages 2135--2135.

\bibitem{horovod}
A.~Sergeev and M.~D. Balso.
\newblock {Horovod: Fast and Easy Distributed Deep Learning in TensorFlow}.
\newblock {\em CoRR}, arXiv/1802.05799, 2018.

\bibitem{memory-mgmt:ipdps-2019}
S.~B. Shriram, A.~Garg, and P.~Kulkarni.
\newblock {Dynamic Memory Management for GPU-based Training of Deep Neural
  Networks}.
\newblock In {\em {IPDPS 2019}}.

\bibitem{dont-decrease-lr-increase-batch}
S.~L. Smith, P.-J. Kindermans, C.~Ying, and Q.~V. Le.
\newblock {Don't Decay the Learning Rate, Increase the Batch Size}.
\newblock In {\em ICLR 2018}.

\bibitem{attention-xml}
R.~You, S.~Dai, Z.~Zhang, H.~Mamitsuka, and S.~Zhu.
\newblock {AttentionXML: Extreme Multi-Label Text Classification with
  Multi-Label Attention Based Recurrent Neural Network}.
\newblock In {\em NeurIPS 2019}.

\bibitem{imagenet-minutes:iclr-2020}
Y.~You, J.~Li, S.~Reddi, J.~Hseu, S.~Kumar, S.~Bhojanapalli, X.~Song,
  J.~Demmel, K.~Keutzer, and C.-J. Hsieh.
\newblock {Large Batch Optimization for Deep Learning: Training BERT in 76
  Minutes}.
\newblock In {\em {ICLR 2020}}.

\bibitem{imagenet-minutes:icpp-2018}
Y.~You, Z.~Zhang, C.-J. Hsieh, J.~Demmel, and K.~Keutzer.
\newblock {ImageNet Training in Minutes}.
\newblock In {\em {ICPP 2018}}.

\bibitem{elastic-sgd}
S.~Zhang, A.~Choromanska, and Y.~LeCun.
\newblock {Deep Learning with Elastic Averaging SGD}.
\newblock In {\em NIPS 2015}, pages 685--693.

\bibitem{zhou2018convergence}
F.~Zhou and G.~Cong.
\newblock {On the Convergence Properties of A K-step Averaging Stochastic
  Gradient Descent Algorithm for Nonconvex Optimization}.
\newblock In {\em IJCAI 2018}.

\bibitem{aws-ec2-gpu}
{Amazon EC2 Instance Types}.
\newblock \url{https://aws.amazon.com/ec2/instance-types/}, 2021.

\bibitem{mxnet-web}
{Apache MXNet}.
\newblock \url{https://mxnet.incubator.apache.org/}, 2021.

\bibitem{hetero-gpu:github}
{HeteroGPU: A Heterogeneity-Aware Multi-GPU Framework for Gradient Descent}.
\newblock \url{https://github.com/YMA33/HeteroGPU}, 2021.

\bibitem{horovod-web}
{Horovod}.
\newblock \url{https://horovod.ai/}, 2021.

\bibitem{cntk-web}
{Microsoft Cognitive Toolkit}.
\newblock \url{https://docs.microsoft.com/en-us/cognitive-toolkit/}, 2021.

\bibitem{nvidia-nccl}
{Nvidia Collective Communication Library (NCCL)}.
\newblock \url{https://developer.nvidia.com/nccl}, 2021.

\bibitem{perlmutter}
{Perlmutter}.
\newblock \url{https://www.nersc.gov/systems/perlmutter/}, 2021.

\bibitem{pytorch-web}
{PyTorch}.
\newblock \url{https://pytorch.org/}, 2021.

\bibitem{slide-code}
{SLIDE}.
\newblock \url{https://github.com/keroro824/HashingDeepLearning}, 2020.

\bibitem{tensorflow-web}
{TensorFlow}.
\newblock \url{https://www.tensorflow.org/}, 2021.

\end{thebibliography}

%%%%%%%%%%%%%%%%%%%%%%%%%%%%%%%%%%%%%%%%%%%%%%%%%%%%%%
\end{document}